\documentclass[epj2,nopacs]{svjour}
\usepackage{graphics}
\usepackage{epsfig}
\usepackage[pagewise,switch]{lineno}
\graphicspath{{figures/}}
\usepackage{subfigure}
\usepackage{multirow}
\usepackage{xspace}

\RequirePackage{ifpdf}
\RequirePackage{url}
\ifpdf
  \RequirePackage[pdftex,colorlinks]{hyperref}
\else
 \RequirePackage{hyperref}
\fi

\def\mm   {\ensuremath{{\rm \,mm}}\xspace}
\def\mum  {\ensuremath{{\,\mu\rm m}}\xspace}
\def\mus        {\ensuremath{\,\mu{\rm s}}\xspace}    
\def\ns         {\ensuremath{{\rm \,ns}}\xspace}      
\def\V   {\ensuremath{\rm \,V}\xspace}
\def\kV   {\ensuremath{\rm \,kV}\xspace}

\begin{document}


\hugehead


\title{Drift Time Measurement in the ATLAS Liquid Argon Electromagnetic 
Calorimeter using Cosmic Muons }

\author{G.~Aad$^{\rm 48}$, B.~Abbott$^{\rm 110}$, J.~Abdallah$^{\rm 11}$,
A.A.~Abdelalim$^{\rm 49}$, A.~Abdesselam$^{\rm 117}$, O.~Abdinov$^{\rm 10}$,
B.~Abi$^{\rm 111}$, M.~Abolins$^{\rm 88}$, H.~Abramowicz$^{\rm 151}$,
H.~Abreu$^{\rm 114}$
% ...
}

\institute{ 
University at Albany, 1400 Washington Ave, Albany, NY 12222, United States of America \and 
University of Alberta, Department of Physics, Centre for Particle Physics, Edmonton, AB T6G 2G7, Canada \and 
Ankara University, Faculty of Sciences, Department of Physics, TR 061000 Tandogan, Ankara, Turkey \and 
LAPP, Universit\'e de Savoie, CNRS/IN2P3, Annecy-le-Vieux, France \and 
Argonne National Laboratory, High Energy Physics Division, 9700 S. Cass Avenue, Argonne IL 60439, United States of America \and 
University of Arizona, Department of Physics, Tucson, AZ 85721, United States of America \and 
The University of Texas at Arlington, Department of Physics, Box 19059, Arlington, TX 76019, United States of America \and 
University of Athens, Nuclear \& Particle Physics, Department of Physics, Panepistimiopouli, Zografou, GR 15771 Athens, Greece \and 
National Technical University of Athens, Physics Department, 9-Iroon Polytechniou, GR 15780 Zografou, Greece \and 
Institute of Physics, Azerbaijan Academy of Sciences, H. Javid Avenue 33, AZ 143 Baku, Azerbaijan \and 
Institut de F\'isica d'Altes Energies, IFAE, Edifici Cn, Universitat Aut\`onoma  de Barcelona,  ES - 08193 Bellaterra (Barcelona), Spain \and 
$^{(a)}$University of Belgrade, Institute of Physics, P.O. Box 57, 11001 Belgrade; Vinca Institute of Nuclear Sciences$^{(b)}$, Mihajla Petrovica Alasa 12-14,
11001 Belgrade, Serbia \and 
University of Bergen, Department for Physics and Technology, Allegaten 55, NO - 5007 Bergen, Norway \and 
Lawrence Berkeley National Laboratory and University of California, Physics Division, MS50B-6227, 1 Cyclotron Road, Berkeley, CA 94720, United States of America
\and Humboldt University, Institute of Physics, Berlin, Newtonstr. 15, D-12489 Berlin, Germany \and 
University of Bern,
Albert Einstein Center for Fundamental Physics,
Laboratory for High Energy Physics, Sidlerstrasse 5, CH - 3012 Bern, Switzerland \and 
University of Birmingham, School of Physics and Astronomy, Edgbaston, Birmingham B15 2TT, United Kingdom 
\and Bogazici University, Faculty of Sciences, Department of Physics, TR - 80815 Bebek-Istanbul, Turkey 
\and INFN Sezione di Bologna$^{(a)}$; Universit\`a  di Bologna, Dipartimento di Fisica$^{(b)}$, viale C. Berti Pichat, 6/2, IT - 40127 Bologna, Italy 
\and University of Bonn, Physikalisches Institut, Nussallee 12, D - 53115 Bonn, Germany 
\and Boston University, Department of Physics,  590 Commonwealth Avenue, Boston, MA 02215, United States of America 
\and Brandeis University, Department of Physics, MS057, 415 South Street, Waltham, MA 02454, United States of America 
\and Universidade Federal do Rio De Janeiro, Instituto de Fisica$^{(a)}$, Caixa Postal 68528, Ilha do Fundao, BR - 21945-970 Rio de Janeiro; $^{(b)}$Universidade de Sao Paulo, Instituto de Fisica, R.do Matao Trav. R.187, Sao Paulo - SP, 05508 - 900, Brazil 
\and Brookhaven National Laboratory, Physics Department, Bldg. 510A, Upton, NY 11973, United States of America 
\and National Institute of Physics and Nuclear Engineering$^{(a)}$, Bucharest-Magurele, Str. Atomistilor 407,  P.O. Box MG-6, R-077125, Romania; $^{(b)}$University Politehnica Bucharest, Rectorat - AN 001, 313 Splaiul Independentei, sector 6, 060042 Bucuresti; $^{(c)}$West University in Timisoara, Bd. Vasile Parvan 4, Timisoara, Romania 
\and Universidad de Buenos Aires, FCEyN, Dto. Fisica, Pab I - C. Universitaria, 1428 Buenos Aires, Argentina 
\and University of Cambridge, Cavendish Laboratory, J J Thomson Avenue, Cambridge CB3 0HE, United Kingdom 
\and Carleton University, Department of Physics, 1125 Colonel By Drive,  Ottawa ON  K1S 5B6, Canada 
\and CERN, CH - 1211 Geneva 23, Switzerland 
\and University of Chicago, Enrico Fermi Institute, 5640 S. Ellis Avenue, Chicago, IL 60637, United States of America 
\and Pontificia Universidad Cat\'olica de Chile, Facultad de Fisica, Departamento de Fisica$^{(a)}$, Avda. Vicuna Mackenna 4860, San Joaquin, Santiago; Universidad T\'ecnica Federico Santa Mar\'ia, Departamento de F\'isica$^{(b)}$, Avda. Esp\~ana 1680, Casilla 110-V,  Valpara\'iso, Chile 
\and Institute of HEP, Chinese Academy of Sciences, P.O. Box 918, CN-100049 Beijing; USTC, Department of Modern Physics, Hefei, CN-230026 Anhui; Nanjing University, Department of Physics,  CN-210093 Nanjing; Shandong University, HEP Group, CN-250100 Shadong, China 
\and Laboratoire de Physique Corpusculaire, CNRS-IN2P3, Universit\'e Blaise Pascal, FR - 63177 Aubiere Cedex, France 
\and Columbia University, Nevis Laboratory, 136 So. Broadway, Irvington, NY 10533, United States of America 
\and University of Copenhagen, Niels Bohr Institute, Blegdamsvej 17, DK - 2100 Kobenhavn 0, Denmark 
\and INFN Gruppo Collegato di Cosenza$^{(a)}$; Universit\`a della Calabria, Dipartimento di Fisica$^{(b)}$, IT-87036 Arcavacata di Rende, Italy 
\and Faculty of Physics and Applied Computer Science of the AGH-University of Science and Technology, (FPACS, AGH-UST), al. Mickiewicza 30, PL-30059 Cracow, Poland 
\and The Henryk Niewodniczanski Institute of Nuclear Physics, Polish Academy of Sciences, ul. Radzikowskiego 152, PL - 31342 Krakow, Poland 
\and Southern Methodist University, Physics Department, 106 Fondren Science Building, Dallas, TX 75275-0175, United States of America \and University of Texas at Dallas, 800 West Campbell Road, Richardson, TX 75080-3021, United States of America 
\and DESY, Notkestr. 85, D-22603 Hamburg , Germany and Platanenallee 6, D-15738 Zeuthen, Germany 
\and TU Dortmund, Experimentelle Physik IV, DE - 44221 Dortmund, Germany 
\and Technical University Dresden, Institut fuer Kern- und Teilchenphysik, Zellescher Weg 19, D-01069 Dresden, Germany 
\and Duke University, Department of Physics, Durham, NC 27708, United States of America 
\and University of Edinburgh, School of Physics \& Astronomy, James Clerk Maxwell Building, The Kings Buildings, Mayfield Road, Edinburgh EH9 3JZ, United Kingdom 
\and Fachhochschule Wiener Neustadt; Johannes Gutenbergstrasse 3 AT - 2700 Wiener Neustadt, Austria 
\and INFN Laboratori Nazionali di Frascati, via Enrico Fermi 40, IT-00044 Frascati, Italy 
\and Albert-Ludwigs-Universit\"{a}t, Fakult\"{a}t f\"{u}r Mathematik und Physik, Hermann-Herder Str. 3, D - 79104 Freiburg i.Br., Germany 
\and Universit\'e de Gen\`eve, Section de Physique, 24 rue Ernest Ansermet, CH - 1211 Geneve 4, Switzerland 
\and INFN Sezione di Genova$^{(a)}$; Universit\`a  di Genova, Dipartimento di Fisica$^{(b)}$, via Dodecaneso 33, IT - 16146 Genova, Italy 
\and Institute of Physics of the Georgian Academy of Sciences, 6 Tamarashvili St., GE - 380077 Tbilisi; Tbilisi State University, HEP Institute, University St. 9, GE - 380086 Tbilisi, Georgia 
\and Justus-Liebig-Universitaet Giessen, II Physikalisches Institut, Heinrich-Buff Ring 16,  D-35392 Giessen, Germany 
\and University of Glasgow, Department of Physics and Astronomy, Glasgow G12 8QQ, United Kingdom 
\and Georg-August-Universitat, II. Physikalisches Institut, Friedrich-Hund Platz 1, D-37077 Goettingen, Germany 
\and Laboratoire de Physique Subatomique et de Cosmologie, CNRS/IN2P3, Universit\'e Joseph Fourier, INPG, 53 avenue des Martyrs, FR - 38026 Grenoble Cedex, France 
\and Hampton University, Department of Physics, Hampton, VA 23668, United States of America 
\and Harvard University, Laboratory for Particle Physics and Cosmology, 18 Hammond Street, Cambridge, MA 02138, United States of America 
\and Ruprecht-Karls-Universitaet Heidelberg, Kirchhoff-Institut fuer Physik$^{(a)}$, Im Neuenheimer Feld 227, D-69120 Heidelberg; $^{(b)}$Physikalisches Institut, Philosophenweg 12, D-69120 Heidelberg; ZITI Ruprecht-Karls-University Heidelberg$^{(c)}$, Lehrstuhl fuer Informatik V, B6, 23-29, DE - 68131 Mannheim, Germany 
\and Hiroshima University, Faculty of Science, 1-3-1 Kagamiyama, Higashihiroshima-shi, JP - Hiroshima 739-8526, Japan 
\and Hiroshima Institute of Technology, Faculty of Applied Information Science, 2-1-1 Miyake Saeki-ku, Hiroshima-shi, JP - Hiroshima 731-5193, Japan 
\and Indiana University, Department of Physics,  Swain Hall West 117, Bloomington, IN 47405-7105, United States of America 
\and Institut fuer Astro- und Teilchenphysik, Technikerstrasse 25, A - 6020 Innsbruck, Austria 
\and University of Iowa, 203 Van Allen Hall, Iowa City, IA 52242-1479, United States of America 
\and Iowa State University, Department of Physics and Astronomy, Ames High Energy Physics Group,  Ames, IA 50011-3160, United States of America 
\and Joint Institute for Nuclear Research, JINR Dubna, RU - 141 980 Moscow Region, Russia 
\and KEK, High Energy Accelerator Research Organization, 1-1 Oho, Tsukuba-shi, Ibaraki-ken 305-0801, Japan 
\and Kobe University, Graduate School of Science, 1-1 Rokkodai-cho, Nada-ku, JP Kobe 657-8501, Japan 
\and Kyoto University, Faculty of Science, Oiwake-cho, Kitashirakawa, Sakyou-ku, Kyoto-shi, JP - Kyoto 606-8502, Japan 
\and Kyoto University of Education, 1 Fukakusa, Fujimori, fushimi-ku, Kyoto-shi, JP - Kyoto 612-8522, Japan 
\and Universidad Nacional de La Plata, FCE, Departamento de F\'{i}sica, IFLP (CONICET-UNLP),   C.C. 67,  1900 La Plata, Argentina 
\and Lancaster University, Physics Department, Lancaster LA1 4YB, United Kingdom 
\and INFN Sezione di Lecce$^{(a)}$; Universit\`a  del Salento, Dipartimento di Fisica$^{(b)}$Via Arnesano IT - 73100 Lecce, Italy 
\and University of Liverpool, Oliver Lodge Laboratory, P.O. Box 147, Oxford Street,  Liverpool L69 3BX, United Kingdom 
\and Jo\v{z}ef Stefan Institute and University of Ljubljana, Department  of Physics, SI-1000 Ljubljana, Slovenia 
\and Queen Mary University of London, Department of Physics, Mile End Road, London E1 4NS, United Kingdom \and
 Royal Holloway, University of London, Department of Physics, Egham Hill, Egham, Surrey TW20 0EX, United Kingdom \and
 University College London, Department of Physics and Astronomy, Gower Street, London WC1E 6BT, United Kingdom \and
 Laboratoire de Physique Nucl\'eaire et de Hautes Energies, Universit\'e Pierre et Marie Curie (Paris 6), Universit\'e Denis Diderot (Paris-7), CNRS/IN2P3, Tour 33, 4 place Jussieu, FR - 75252 Paris Cedex 05, France \and
 Lunds universitet, Naturvetenskapliga fakulteten, Fysiska institutionen, Box 118, SE - 221 00 Lund, Sweden \and
 Universidad Autonoma de Madrid, Facultad de Ciencias, Departamento de Fisica Teorica, ES - 28049 Madrid, Spain \and
 Universitaet Mainz, Institut fuer Physik, Staudinger Weg 7, DE - 55099 Mainz, Germany \and
 University of Manchester, School of Physics and Astronomy, Manchester M13 9PL, United Kingdom \and
 CPPM, Aix-Marseille Universit\'e, CNRS/IN2P3, Marseille, France \and
 University of Massachusetts, Department of Physics, 710 North Pleasant Street, Amherst, MA 01003, United States of America \and
 McGill University, High Energy Physics Group, 3600 University Street, Montreal, Quebec H3A 2T8, Canada \and
 University of Melbourne, School of Physics, AU - Parkville, Victoria 3010, Australia \and
 The University of Michigan, Department of Physics, 2477 Randall Laboratory, 500 East University, Ann Arbor, MI 48109-1120, United States of America \and
 Michigan State University, Department of Physics and Astronomy, High Energy Physics Group, East Lansing, MI 48824-2320, United States of America \and
 INFN Sezione di Milano$^{(a)}$; Universit\`a  di Milano, Dipartimento di Fisica$^{(b)}$, via Celoria 16, IT - 20133 Milano, Italy \and
 B.I. Stepanov Institute of Physics, National Academy of Sciences of Belarus, Independence Avenue 68, Minsk 220072, Republic of Belarus \and
 National Scientific \& Educational Centre for Particle \& High Energy Physics, NC PHEP BSU, M. Bogdanovich St. 153, Minsk 220040, Republic of Belarus \and
 Massachusetts Institute of Technology, Department of Physics, Room 24-516, Cambridge, MA 02139, United States of America \and
 University of Montreal, Group of Particle Physics, C.P. 6128, Succursale Centre-Ville, Montreal, Quebec, H3C 3J7  , Canada \and
 P.N. Lebedev Institute of Physics, Academy of Sciences, Leninsky pr. 53, RU - 117 924 Moscow, Russia \and
 Institute for Theoretical and Experimental Physics (ITEP), B. Cheremushkinskaya ul. 25, RU 117 218 Moscow, Russia \and
 Moscow Engineering \& Physics Institute (MEPhI), Kashirskoe Shosse 31, RU - 115409 Moscow, Russia \and
 Lomonosov Moscow State University Skobeltsyn Institute of Nuclear Physics (MSU SINP), 1(2), Leninskie gory, GSP-1, Moscow 119991 Russian Federation , Russia \and
 Ludwig-Maximilians-Universit\"at M\"unchen, Fakult\"at f\"ur Physik, Am Coulombwall 1,  DE - 85748 Garching, Germany \and
 Max-Planck-Institut f\"ur Physik, (Werner-Heisenberg-Institut), F\"ohringer Ring 6, 80805 M\"unchen, Germany \and
 Nagasaki Institute of Applied Science, 536 Aba-machi, JP Nagasaki 851-0193, Japan \and
 Nagoya University, Graduate School of Science, Furo-Cho, Chikusa-ku, Nagoya, 464-8602, Japan \and
 INFN Sezione di Napoli$^{(a)}$; Universit\`a  di Napoli, Dipartimento di Scienze Fisiche$^{(b)}$, Complesso Universitario di Monte Sant'Angelo, via Cinthia, IT - 80126 Napoli, Italy \and
  University of New Mexico, Department of Physics and Astronomy, MSC07 4220, Albuquerque, NM 87131 USA, United States of America \and
 Radboud University Nijmegen/NIKHEF, Department of Experimental High Energy Physics, Toernooiveld 1, NL - 6525 ED Nijmegen , Netherlands \and
 Nikhef National Institute for Subatomic Physics, and University of Amsterdam, Science Park 105, 1098 XG Amsterdam, Netherlands \and
 Budker Institute of Nuclear Physics (BINP), RU - Novosibirsk 630 090, Russia \and
 New York University, Department of Physics, 4 Washington Place, New York NY 10003, USA, United States of America \and
 Ohio State University, 191 West Woodruff Ave, Columbus, OH 43210-1117, United States of America \and
 Okayama University, Faculty of Science, Tsushimanaka 3-1-1, Okayama 700-8530, Japan \and
 University of Oklahoma, Homer L. Dodge Department of Physics and Astronomy, 440 West Brooks, Room 100, Norman, OK 73019-0225, United States of America \and
 Oklahoma State University, Department of Physics, 145 Physical Sciences Building, Stillwater, OK 74078-3072, United States of America \and
 Palack\'y University, 17.listopadu 50a,  772 07  Olomouc, Czech Republic \and
 University of Oregon, Center for High Energy Physics, Eugene, OR 97403-1274, United States of America \and
 LAL, Univ. Paris-Sud, IN2P3/CNRS, Orsay, France \and
 Osaka University, Graduate School of Science, Machikaneyama-machi 1-1, Toyonaka, Osaka 560-0043, Japan \and
 University of Oslo, Department of Physics, P.O. Box 1048,  Blindern, NO - 0316 Oslo 3, Norway \and
 Oxford University, Department of Physics, Denys Wilkinson Building, Keble Road, Oxford OX1 3RH, United Kingdom \and
 INFN Sezione di Pavia$^{(a)}$; Universit\`a  di Pavia, Dipartimento di Fisica Nucleare e Teorica$^{(b)}$, Via Bassi 6, IT-27100 Pavia, Italy \and
 University of Pennsylvania, Department of Physics, High Energy Physics Group, 209 S. 33rd Street, Philadelphia, PA 19104, United States of America \and
 Petersburg Nuclear Physics Institute, RU - 188 300 Gatchina, Russia \and
 INFN Sezione di Pisa$^{(a)}$; Universit\`a   di Pisa, Dipartimento di Fisica E. Fermi$^{(b)}$, Largo B. Pontecorvo 3, IT - 56127 Pisa, Italy \and
 University of Pittsburgh, Department of Physics and Astronomy, 3941 O'Hara Street, Pittsburgh, PA 15260, United States of America \and
 $^{(a)}$Universidad de Granada, Departamento de Fisica Teorica y del Cosmos and CAFPE, E-18071 Granada; Laboratorio de Instrumentacao e Fisica Experimental de Particulas - LIP$^{(b)}$, Avenida Elias Garcia 14-1, PT - 1000-149 Lisboa, Portugal \and
 Institute of Physics, Academy of Sciences of the Czech Republic, Na Slovance 2, CZ - 18221 Praha 8, Czech Republic \and
 Charles University in Prague, Faculty of Mathematics and Physics, Institute of Particle and Nuclear Physics, V Holesovickach 2, CZ - 18000 Praha 8, Czech Republic \and
 Czech Technical University in Prague, Zikova 4, CZ - 166 35 Praha 6, Czech Republic \and
 State Research Center Institute for High Energy Physics, Moscow Region, 142281, Protvino, Pobeda street, 1, Russia \and
 Rutherford Appleton Laboratory, Science and Technology Facilities Council, Harwell Science and Innovation Campus, Didcot OX11 0QX, United Kingdom \and
 University of Regina, Physics Department, Canada \and
 Ritsumeikan University, Noji Higashi 1 chome 1-1, JP - Kusatsu, Shiga 525-8577, Japan \and
 INFN Sezione di Roma I$^{(a)}$; Universit\`a  La Sapienza, Dipartimento di Fisica$^{(b)}$, Piazzale A. Moro 2, IT- 00185 Roma, Italy \and
 INFN Sezione di Roma Tor Vergata$^{(a)}$; Universit\`a di Roma Tor Vergata, Dipartimento di Fisica$^{(b)}$ , via della Ricerca Scientifica, IT-00133 Roma, Italy \and
 INFN Sezione di  Roma Tre$^{(a)}$; Universit\`a Roma Tre, Dipartimento di Fisica$^{(b)}$, via della Vasca Navale 84, IT-00146  Roma, Italy \and
 Universit\'e Hassan II, Facult\'e des Sciences Ain Chock$^{(a)}$, B.P. 5366, MA - Casablanca; Centre National de l'Energie des Sciences Techniques Nucleaires (CNESTEN)$^{(b)}$, B.P. 1382 R.P. 10001 Rabat 10001; Universit\'e Mohamed Premier$^{(c)}$, LPTPM, Facult\'e des Sciences, B.P.717. Bd. Mohamed VI, 60000, Oujda ; Universit\'e Mohammed V, Facult\'e des Sciences$^{(d)}$, LPNR, BP 1014, 10000 Rabat, Morocco \and
 CEA, DSM/IRFU, Centre d'Etudes de Saclay, FR - 91191 Gif-sur-Yvette, France \and
 University of California Santa Cruz, Santa Cruz Institute for Particle Physics (SCIPP), Santa Cruz, CA 95064, United States of America \and
 University of Washington, Seattle, Department of Physics, Box 351560, Seattle, WA 98195-1560, United States of America \and
University of Sheffield, Department of Physics \& Astronomy, Hounsfield Road, Sheffield S3 7RH, United Kingdom \and
 Shinshu University, Department of Physics, Faculty of Science, 3-1-1 Asahi, Matsumoto-shi, JP - Nagano 390-8621, Japan \and
 Universitaet Siegen, Fachbereich Physik, D 57068 Siegen, Germany \and
 Simon Fraser University, Department of Physics, 8888 University Drive, CA - Burnaby, BC V5A 1S6, Canada \and
 SLAC National Accelerator Laboratory, Stanford, California 94309, United States of America \and
 Comenius University, Faculty of Mathematics, Physics \& Informatics, Mlynska dolina F2, SK - 84248 Bratislava; Institute of Experimental Physics of the Slovak Academy of Sciences, Dept. of Subnuclear Physics, Watsonova 47, SK - 04353 Kosice, Slovak Republic \and
 Stockholm University, Department of Physics, AlbaNova, SE - 106 91 Stockholm, Sweden \and
 Royal Institute of Technology (KTH), Physics Department, SE - 106 91 Stockholm, Sweden \and
 Stony Brook University, Department of Physics and Astronomy, Nicolls Road, Stony Brook, NY 11794-3800, United States of America \and
 University of Sussex, Department of Physics and Astronomy
Pevensey 2 Building, Falmer, Brighton BN1 9QH, United Kingdom \and
 University of Sydney, School of Physics, AU - Sydney NSW 2006, Australia \and
 Insitute of Physics, Academia Sinica, TW - Taipei 11529, Taiwan \and
 Technion, Israel Inst. of Technology, Department of Physics, Technion City, IL - Haifa 32000, Israel \and
 Tel Aviv University, Raymond and Beverly Sackler School of Physics and Astronomy, Ramat Aviv, IL - Tel Aviv 69978, Israel
% ...
}

\date{Received: date / Revised version: date}
\abstract{
The ionization signals in the liquid argon of the ATLAS electromagnetic calorimeter are
studied in detail using cosmic muons. In particular, the drift time of the ionization
electrons is measured and used to assess the intrinsic uniformity of the 
calorimeter gaps and estimate its impact on the constant term of the energy
resolution. The drift times of electrons in the cells of the second layer 
of the calorimeter are uniform at the level of $1.3 \ \%$ in the barrel and 
$2.8 \ \%$ in the endcaps. This leads to an estimated
contribution to the constant term of $(0.29^{+0.05}_{-0.04}) \ \%$ in the barrel and $(0.54^{+0.06}_{-0.04}) \ \%$ in the endcaps.
The same data are used to measure the drift velocity of ionization electrons in liquid argon,
which is found to be $4.61 \pm 0.07 \rm \ mm/\mu s$ at $88.5 \rm \ K$ and $1 \rm \ kV/mm$.
}

\maketitle



\section{Introduction}
\label{sect:intro}

The ATLAS liquid argon (LAr) calorimeter~\cite{detectorpaper} is composed of sampling 
detectors with full azimuthal\footnote{The azimuthal angle $\phi$ is measured in the 
plane transverse to the beam axis. Positive $\phi$ is in the up direction. 
The pseudorapidity is defined as $\eta = -\ln(\tan(\theta/2))$, where $\theta$ is 
the polar angle from the beam axis. Positive $\eta$ is for the proton beam circulating 
anticlockwise.} symmetry and is housed in one barrel and two endcap cryostats. A highly 
granular electromagnetic (EM) calorimeter with accordion--shaped electrodes and lead 
absorbers covers the pseudorapidity range $|\eta|<3.2$, and contains a barrel
part ($|\eta|<1.475$)~\cite{construction} made of two half-barrels joined at $\eta = 0$ 
and two endcap parts ($1.375<|\eta|<3.2$)~\cite{construction_Endcap}. Each section is 
segmented in depth in three layers (denoted as layer 1,2,3). For $|\eta|<1.8$, a 
presampler (PS)~\cite{presampler,construction_Endcap}, installed in the cryostat in 
front of the EM calorimeter, provides a measurement of the energy lost upstream.
 
\begin{figure}
  \centering
  \includegraphics[width=0.9\columnwidth]{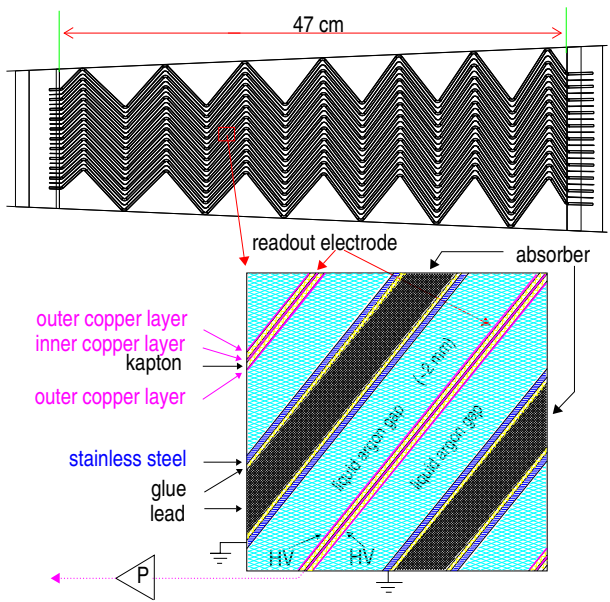}
 \caption{Accordion structure of the barrel. The top figure is a 
 view of a small sector of the barrel calorimeter in a plane transverse 
 to the LHC beams. Honeycomb spacers, in the liquid argon gap, position 
 the electrodes between the lead absorber plates.}
  \label{F:accordion}
\end{figure}

The EM calorimeter plays a crucial role during the operation of the LHC, since physics 
channels involving electrons and photons in the final state form a crucial part of the 
ATLAS physics program. Achieving the required precision and discovery reach places 
stringent requirements on the performance of the calorimeter. The uniformity of the 
calorimeter response over a large acceptance is particularly important for the overall 
resolution. This drives several design choices for the calorimeter: lead-liquid argon 
calorimetry provides a good energy resolution and homogeneity even in the presence 
of strong radiation; the accordion geometry (see Figure~\ref{F:accordion}) avoids readout 
cracks between calorimeter modules, thus also providing good uniformity.

In order to equalize the gains of different calorimeter channels, a calibration procedure 
involving electronic charge injection is used. This is however not sensitive to intrinsic 
characteristics of the ionization gaps in the liquid argon system, such as variations in 
gap sizes and LAr temperature changes. Such non-uniformities can be measured from the ionization 
signals created by charged particles. The calorimeter energy response to this ionization is 
not the best quantity for this purpose, because it requires a knowledge of the energy of the 
incoming particle. However the electron drift time in LAr, which can be obtained from the signal 
pulse shape resulting from ionizing particles that deposit sufficient energy above the intrinsic 
noise level in a calorimeter cell, is a powerful monitoring tool. As explained in 
Section~\ref{sect:ionization}, the drift time is also about four times more sensitive to changes 
in the LAr gap size than is the energy response. Cosmic muons have been used to this end as part 
of the calorimeter commissioning before the LHC start-up.

The EM calorimeter installation in the ATLAS cavern was completed at the end of 2006.
Before LHC start-up, the main challenge was to commission the associated electronics and automate 
all of the calibration steps for the full $173,312$ channels. Cosmic muon data have been taken 
regularly for commissioning purposes since 2006. At the end of the summer and during autumn of
2008 stable cosmic muon runs were taken with the detector fully operational and using various 
trigger menus. In normal data taking only $5$ samples around the pulse peak at $25 \rm \ ns$ 
intervals are taken, but in order to accurately measure the drift time $32$ samples are needed.
The pulse height is also relevant, since larger pulses are less affected by electronic noise. 
A summary of the detector performance obtained from calibration data, cosmic muons and beam 
splash events is detailed in~\cite{readiness}.

Measurements of the drift time ($T_{drift}$) in the ATLAS EM calorimeter using cosmic muon data 
are presented in this paper. These drift times, which are independent of the amplitude of the 
pulses used for their determination, can be compared from one calorimeter region to another, 
and thus allow a measurement of the uniformity of the calorimeter. 


\section{Ionization signal in the calorimeter}
\label{sect:ionization}

The current resulting from the passage of a charged particle through a liquid 
argon gap has the typical ionization-chamber triangular shape, with a short rise time
(smaller than $1 \rm \ ns$) which is neglected in the rest of this note,
followed by a linear decay for the duration of the
maximum drift time

\begin{equation}
T_{drift}=w_{gap}/V_{drift},
\label{eq:pure_vdrift}
\end{equation}

\noindent where $w_{gap}$ is the LAr gap width and $V_{drift}$ the electron drift velocity
\cite{atlascalo}. The ionization current, $I$, is then modeled as:

\begin{equation}
  I(t;I_0,T_{drift}) = I_0 \left( 1-\frac{t}{T_{drift}} \right) \hspace{0.1cm} {\rm for} \ \ 0<t<T_{drift}
  \label{eq:Triangle}
\end{equation}

\noindent where $I_{0}$ is the current at $t=0$.
The peak current amplitude $I_0 = \rho \cdot V_{drift}$
is proportional to the drift velocity and to the negative linear charge density $\rho$
along the direction perpendicular to the readout electrode, which varies with the lead 
thickness~\footnote{If the LAr gap increases (as in the endcap) 
$\rho$ increases slightly on average due to showering in LAr. 
This is accounted for using detector simulation.}.
Since the determination of the energy is based on the
measurement of $I_0$, it is crucial to be able to precisely evaluate and monitor $V_{drift}$. 
While the LAr gap thickness is mechanically constrained, the drift velocity depends on
the actual conditions of the detector: the LAr temperature and density, and
the local high voltage. Uniform response in a calorimeter with constant lead
thickness requires uniform drift velocity in the gaps.

At this point it is appropriate to recall that each 
liquid argon electronic cell is built out of several gaps connected 
in parallel: for layers 2 and 3, there are 4(3) double-gaps in parallel in 
the barrel (endcap) respectively; there are four times as many gaps per cell in layer 1, 
given the coarser granularity of the readout in the azimuthal direction~\cite{detectorpaper}. 
The parameters measured represent an average of the  local gaps, 
both in depth along the cell, and in between the gaps forming a cell.

\begin{figure*}[htbp]
\begin{center}

\subfigure[Layer 2 of barrel]{
\label{F:pulse1}
\includegraphics[width=0.8\columnwidth]{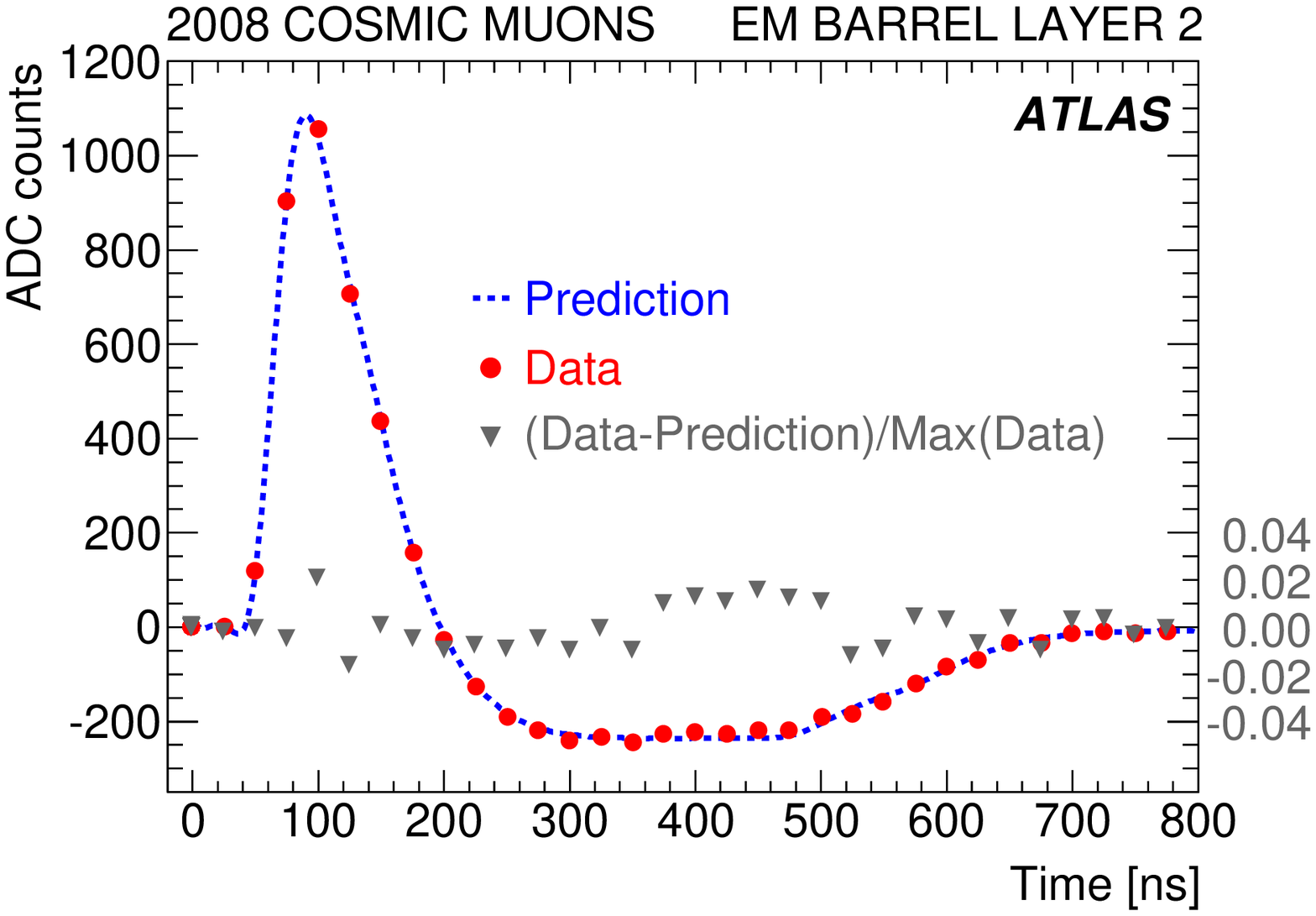}
}
\subfigure[Layer 2 of endcap]{
\label{F:pulse2}
\includegraphics[width=0.8\columnwidth]{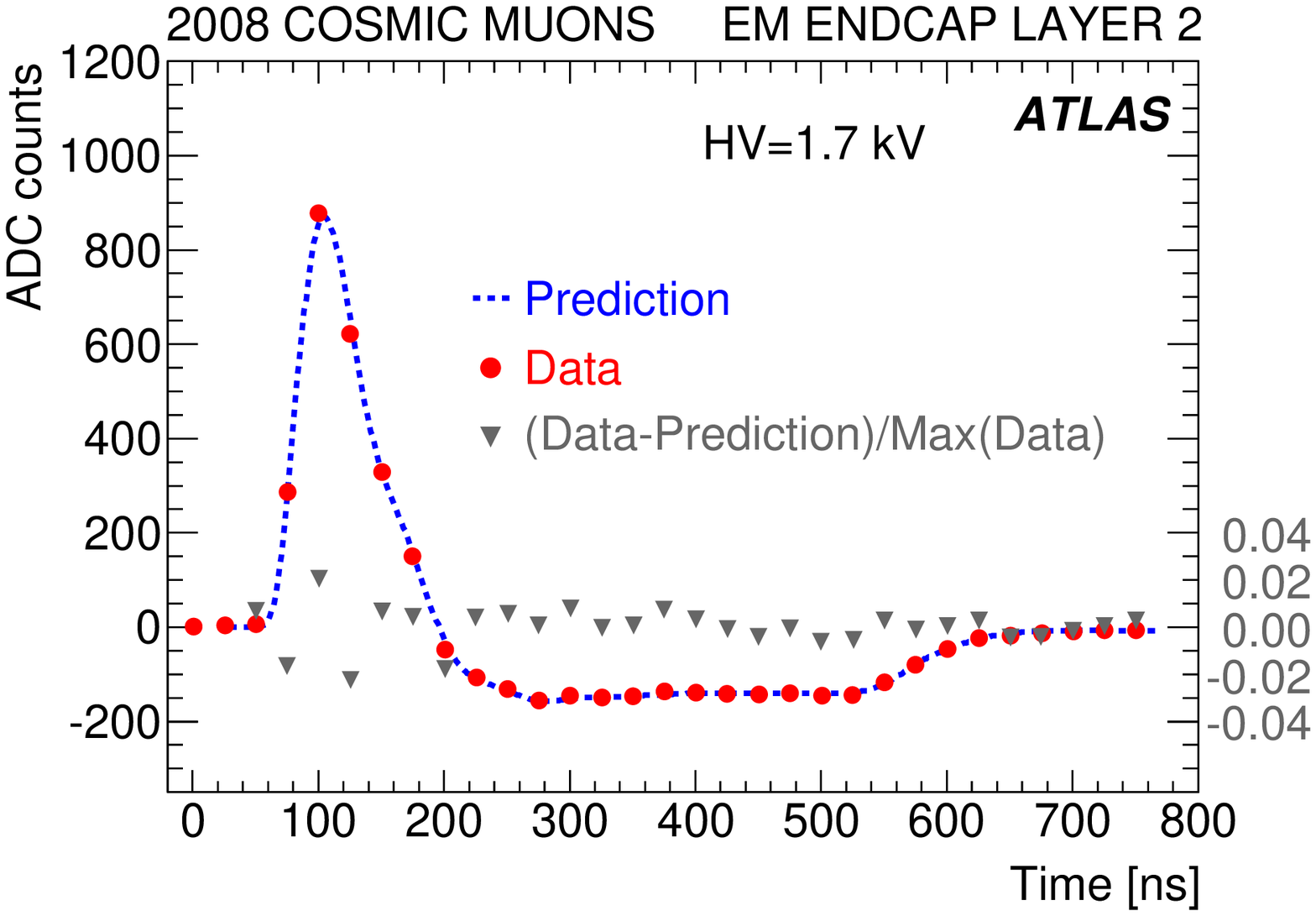}
}
\label{F:pulse_shape}
\caption{Typical single ionization pulse in a cell of layer 2 of the barrel (a)  
and endcap (b)  of the calorimeter. The large red dots show the data samples, 
the small blue dots the prediction and the grey triangles the relative difference 
(data ($S$) - prediction ($g$))/$S_{max}$, on the scale shown on the right side 
of the plot (normalized to the data).}
\end{center}
\end{figure*}

\begin{figure}[htbp]
\begin{center}
\includegraphics[width=0.8\columnwidth]{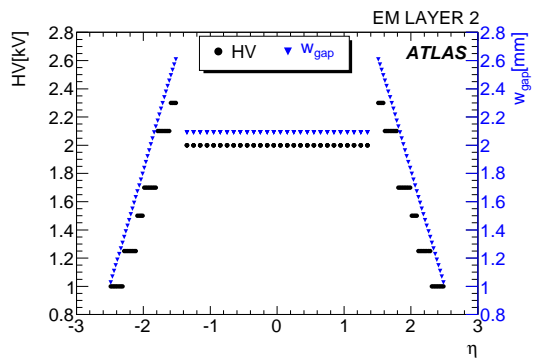}
\caption{Nominal HV (black dots) and nominal gap width $w_{gap}$ (blue triangles) versus 
$\eta$ in the 2nd layer of the EM calorimeter. 
\label{F:hv_distri}}
\end{center}
\end{figure}

At the end of the readout chain the triangular signal is amplified,
shaped and passed through a switched capacitor array which samples
the signal every $25 \rm \ ns$. The shaping function (see Section~\ref{sect:prediction})
includes one integration and two derivatives. Their net effect is to transform
the triangular signal in a positive spike, followed by a flat undershoot, the length
of the undershoot being equal to the drift time. The net area of the pulse, except 
for small fluctuations due to noise, being equal to $0$. Upon Level1 trigger decision, 
the samples are then digitized using a fast-ADC and recorded~\cite{front_ele,back_ele}.
Figure~\ref{F:pulse_shape} shows two typical digitized signal shapes, one for the barrel 
and the other for the endcap. The data samples in each plot correspond to a single 
cosmic muon event in a single cell, and fluctuations of the amplitude in 
each sample due to noise can be observed. The pulses shown pass the analysis criteria 
described in Section~\ref{sect:data}. The prediction is obtained by modeling the readout chain as described in 
Section~\ref{sect:prediction}. In the barrel section, the nominal gap size is 
constant ($2.09 \rm \ mm$); in the endcap the gap size changes with pseudorapidity 
(see Figure~\ref{F:hv_distri}), so that at larger values of $\eta$ smaller gaps lead 
to a shorter pulse undershoot.

\begin{figure}[htbp]
  \centering
  \includegraphics[width=0.7\columnwidth]{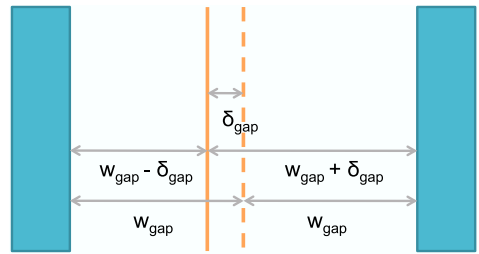}
  \caption{Schematic view of a LAr gap. The nominal position of
    the readout electrode (dashed line) is exactly equidistant from the lead
    absorbers. Any shift with respect to the nominal position (solid line)
    causes an increase of the gap width on one side of the electrode,
    and a decrease on the other side.}
  \label{F:ElectrodeShift}
\end{figure}

\begin{figure}[htbp]
  \centering
  \includegraphics[width=0.8\columnwidth]{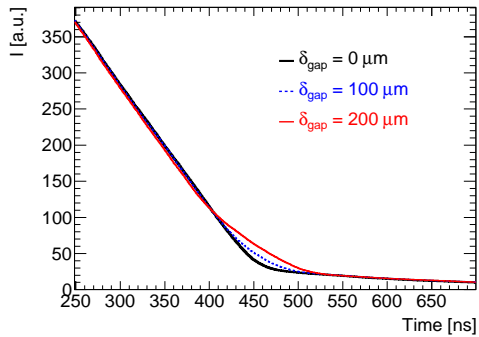}
  \caption{Current as a function of time for a perfect centering of the electrode ($\delta_{gap}=0 \rm \ \mu m$),
  a shift of $\delta_{gap}=100 \rm \ \mu m$ and $\delta_{gap}=200 \rm \ \mu m$.}
  \label{F:current}
\end{figure}

   In the ideal case, an electrode is surrounded by two identical gaps, one on each side (see Figure~\ref{F:ElectrodeShift}).
Any modification of one of the gaps by a relative fraction $x$ will break the symmetry, leading to two different
values of drift time $T_{Di}$ ($i=1,2$) (Equations~\ref{eq:td1} and~\ref{eq:td2}). 
Figure~\ref{F:current} demonstrates this effect by showing the total collected 
current versus time in the case where the electrode is at the nominal position 
($\delta_{gap}=0 \rm \ \mu m$) or shifted by $100 \rm \ \mu m$ and $200 \rm \ \mu m$. 
This affects the rise at the end of the pulse
(between $450$ and $650\rm \ ns$ on Figure~\ref{F:pulse1} for example) which is sensitive to 
changes in the gap size over the charge collection area. 
The variation of the drift time inside the cell arises in part from 
the slight opening of the gaps along the accordion folds (see Figure~\ref{F:accordion}), but 
the bulk of the effect is caused by random or systematic displacements of the electrodes away from the gap center. 
Both effects are parametrized by the shift parameter $\delta_{gap} =x \cdot w_{gap}$. This shift 
parameter is limited to a maximum of $400 \rm \ \mu m$ due to the honeycomb filling the gaps, however, 
some modifications of electrical field lines (like edge effects) can contribute to local enlargements.

\newpage

Beside the gap width, $w_{gap}$, the model of the signal takes into account the electrode 
shift parameter as well as possible variations in high voltage on both sides (neglecting 
in a first description the bend parts.).
The total signal can be expressed as a sum of two triangular signals, one for each side of the gap, each described by
a drift time $T_{D i}$ and peak current $f_{i} \cdot I_{0}$ ($i=1,2$). Since the drift velocity $V_{drift}$ in liquid argon
follows, for the range of electric fields relevant for this study, a power-law dependence on the electric 
field value~\cite{power_law1,power_law2}, with an exponent denoted here by $\alpha$

\begin{equation}
V_{drift} = K \cdot \Big[ \frac{HV}{w_{gap}}\Big]^{\alpha}
\label{eq:vdrift_field}
\end{equation}

\noindent the drift time and peak current fraction are given by:

\begin{eqnarray}
 T_{D1} &=& T_{drift} \ (1- x)^{1+\alpha} \  (HV_1/HV_{nom})^ {-\alpha}  \ , \nonumber \\ 
 f_{1} &=& \frac {f_{nom}}{2} \ (1- x) ^{-\alpha} \ (HV_1/HV_{nom})^ {\alpha}
\label{eq:td1}  \\ 
 T_{D2} &=& T_{drift} \ (1+ x)^{1+\alpha} \  (HV_2/HV_{nom})^ {-\alpha}  \ , \nonumber \\ 
 f_2    &=& \frac {f_{nom}}{2} \ (1+ x) ^{-\alpha} \ (HV_2/HV_{nom})^ {\alpha}
\label{eq:td2}
\end{eqnarray}

\noindent where $T_{drift}$ and $f_{nom}$ ($f_{nom}=1$ when the bend parts are neglected) are respectively the drift time value and the 
fraction of current corresponding to the nominal high voltage $HV_{nom}$, and $HV_i$ 
corresponds to the actual high voltage applied on side $i$. In the barrel the nominal high 
voltage is $2 \rm \ kV$; in the endcap, the high voltage varies with $\eta$ (see Figure~\ref{F:hv_distri})
to cope with the varying gap, ensuring in principle a 
constant drift velocity by keeping the electric field constant. For the high voltage 
distribution, electrodes are grouped by sectors of 
$\Delta \eta \times \Delta \phi = 0.2 \times 0.2$ and for redundancy separated supplies are 
used for each side of the electrodes. While in the vast majority of the sectors the high
voltage has the nominal value, a few of them are operated at lower values, to prevent
accidental sparking or excess noise.


Both in the barrel and in the endcap, the nominal operating field is close to $1 \rm \ kV/mm$.
The range of variation of $x$ (up to typically $20 \ \%$) induces a corresponding variation of the
operating field of $\pm 20 \ \%$. In this reduced range, and for a fixed value of the liquid
argon temperature, $88.5 \rm \ K$, the variation of the drift velocity with the field is well
described~\cite{power_law1,power_law2} by a power law (Equation~\ref{eq:vdrift_field}). Fitting the data published in~\cite{VdvsTF} with such law gives $\alpha_{1} = 0.316 \pm 0.030$.
Additional information was obtained with our own data comparing a group of sectors in the barrel
operated at $1600 \rm \ V$, to the bulk operated at $2000 \rm \ V$. The ratio of the velocity values
obtained, taking into account small position dependence (see Section~\ref{sect:barrel_result}), gives: $\alpha_{2} = 0.295 \pm 0.020$.
Considering these two values, and given the low sensitivity of our results to the
exact value of $\alpha$ (see Section~\ref{sec:systematics}) we decided to use $\alpha=0.3$
with a systematic uncertainty of  $^{+0.04}_{-0.02}$.

\begin{figure*}[htbp]
\begin{center}
\subfigure[MC $T_{drift}$ for endcap]{
\label{F:mc_endcap_td}
\includegraphics[width=0.8\columnwidth]{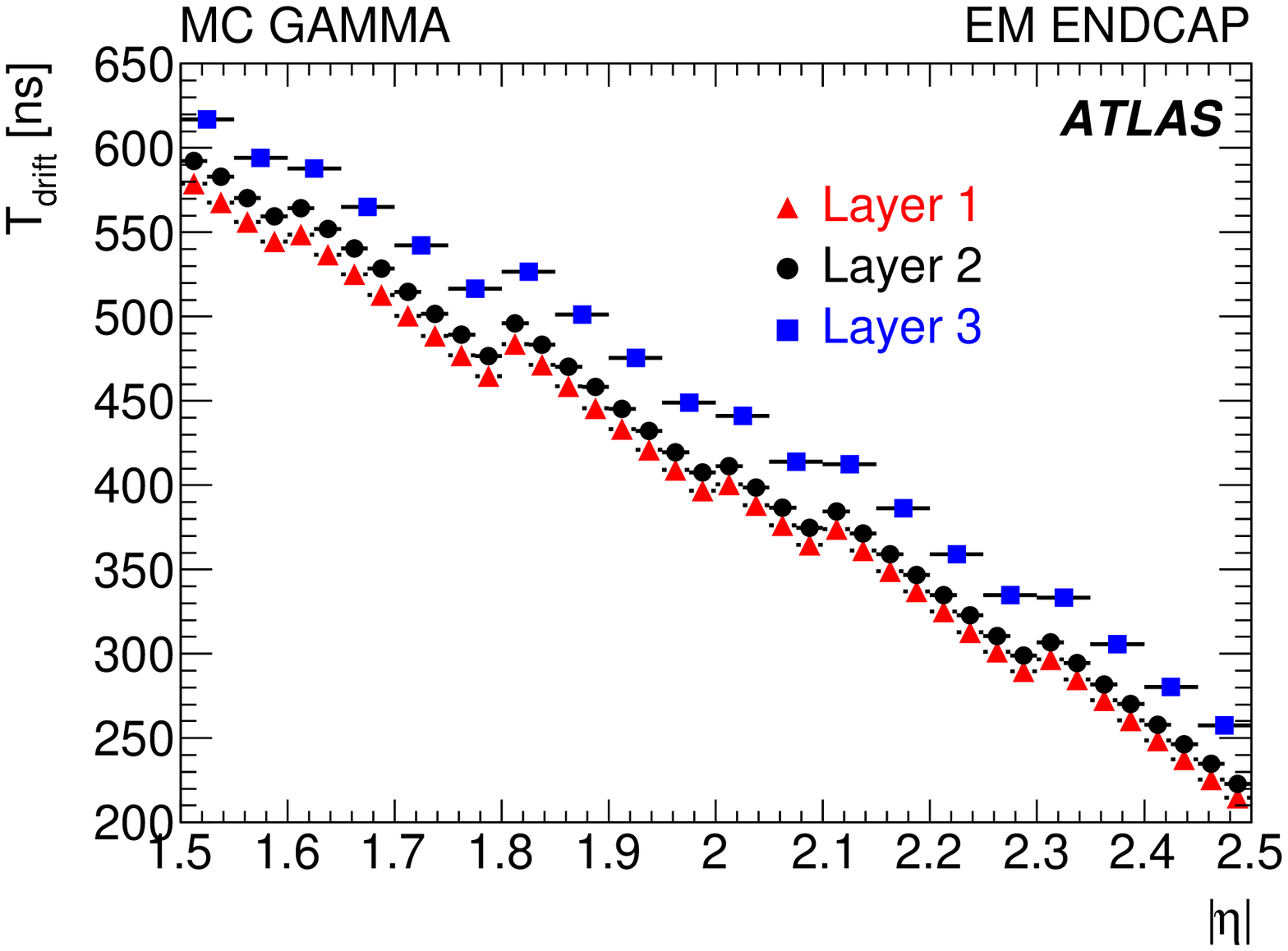}
}
\subfigure[MC $T_{bend}$ for endcap]{
\label{F:mc_endcap_td3}
\includegraphics[width=0.8\columnwidth]{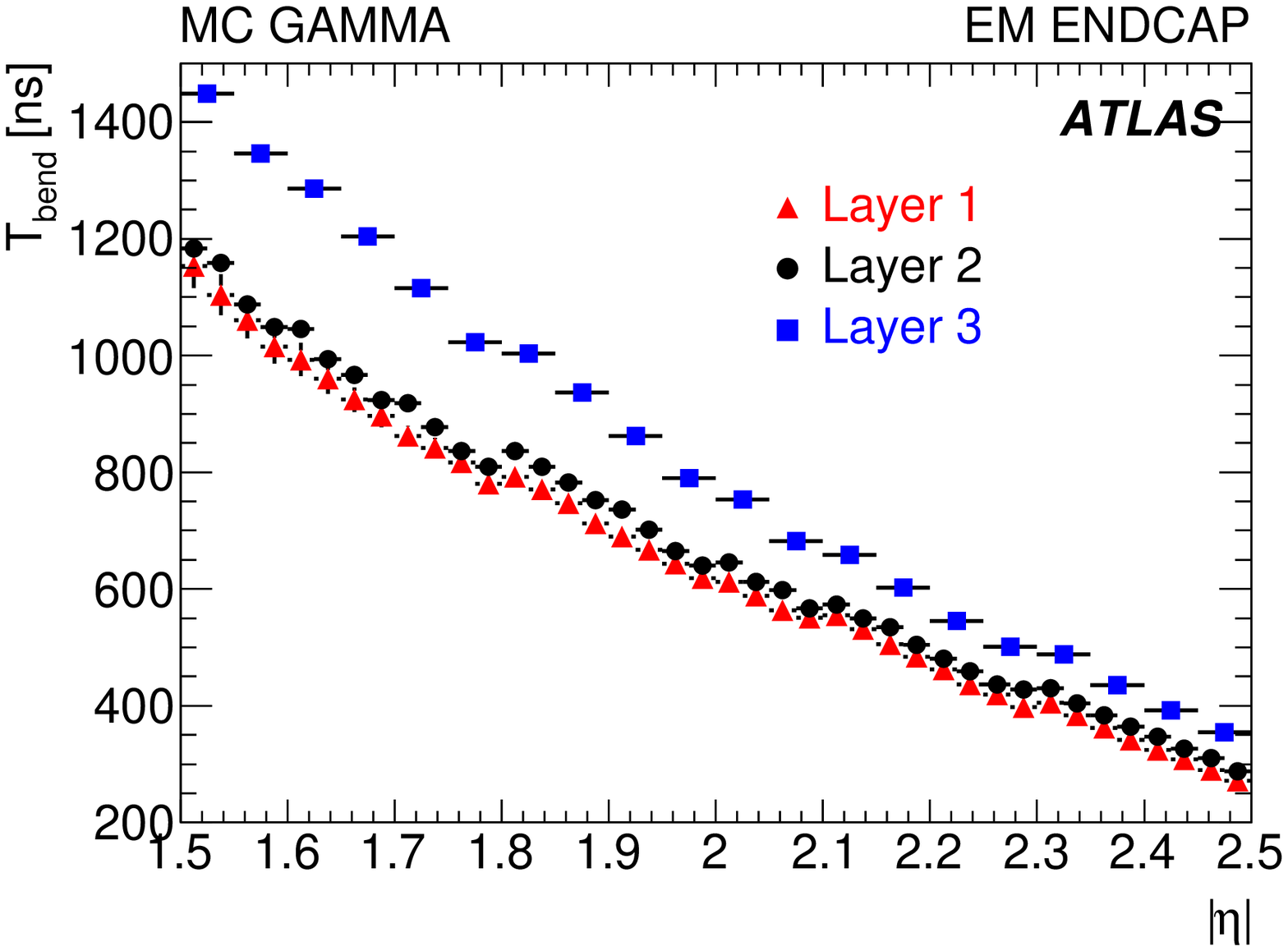}
}
\caption{Monte Carlo simulation for (a) $T_{drift}$ and (b) $T_{bend}$ versus
$\eta$ for the three endcap layers: layer 1 (red triangles), layer 2 (black dots) and layer 3 (blue squares).
\label{F:mc_plots}}
\end{center}
\end{figure*}

In the accordion geometry, the electric field in the bent sections has a lower value than in the flat parts. 
This leads to another contribution to the ionization signal in the form of two smaller triangular signals 
with a longer time constant and smaller $f_{bend}$. The sum of the current
fractions ($f_{nom}+f_{bend}$) must be equal to 1; the main contribution on Figure~\ref{F:current} is related
to the drift time in flat sections, the tail at large time ($t > 500 \rm \ ns$) is due to the larger gap
width in the bent sections of the accordion. These triangular shapes are parametrized (neglecting the electrode shift effect) by

\begin{eqnarray}
 T_{D3} &=& T_{bend} \ (HV_1/HV_{nom})^ {-\alpha} \ ,  \nonumber \\ 
 f_3 &=& \frac {f_{bend}}{2} \ (HV_1/HV_{nom})^ {\alpha} \\
 T_{D4} &=& T_{bend} \ (HV_2/HV_{nom})^ {-\alpha} \ ,  \nonumber \\  
 f_4 &=& \frac {f_{bend}}{2} \ (HV_2/HV_{nom})^ {\alpha}
 \label{eq:td4}
\end{eqnarray}

In the barrel, the $T_{bend}$ and $f_{bend}$ contributions per layer are estimated using the GEANT4 simulation 
of a uniform charge density in the gap.
These values are given in Table~\ref{tab:bend} for layers 1 to 3 (there are no bent
sections in the presampler).

\begin{table}[htbp]
  \centering
  \begin{tabular}{llll}
    \hline \hline
    & Layer & $T_{bend}$ ($\rm \ ns$) & $f_{bend}$ (\%) \\
    \hline
    & Layer 1      &  820.   &  4.9 \\
    & Layer 2     &  898.   &  7.1 \\
    & Layer 3       &  941.   &  8.5 \\
    \hline \hline
  \end{tabular} 
  \caption{\label{tab:bend}  $T_{bend}$ and $f_{bend}$ values for the
different layers in the barrel.}
\end{table}

In the endcaps, for practical reasons 
a different simulation was used, MC GAMMA, where $10 \rm \ GeV$
electromagnetic showers have been simulated to 
predict the drift time $T_{drift}$ and to estimate $T_{bend}$ and 
$f_{bend}$. A photon simulation was chosen since the signals relevant 
to this study originate from electromagnetic showers produced by 
cosmic muons. The simulated photons were generated with a flight 
direction originating from the ATLAS Interaction Point. This differs 
from cosmic muons which cross the calorimeter in a quasi-vertical 
direction. Both $T_{drift}$ and $T_{bend}$ are plotted in  
Figure~\ref{F:mc_plots} as a function of pseudorapidity for the three 
layers. These quantities are obtained from the distribution of the 
local drift time where the contributions from straight and bent sections 
of the accordion are clearly distinguished. Figure~\ref{F:mc_plots} shows 
that both quantities decrease with increasing $\eta$, following the 
reduction of the gap size. The difference observed between the layers 
is due to the depth variation of the gap size: the gap grows continuously 
from layer 1 to 3 due to the projective geometry of the cells. The values 
for layer 2 lie closer to those of layer 1. This is explained by the fact 
that at the energy of the simulated showers ($10 \rm \ GeV$), the shower 
maximum is closer to layer 1 than to layer 3. The current fraction $f_{bend}$ 
is also estimated from the simulation for every $\eta$ cell, with 
values ranging from $5 \%$ to $20 \%$ depending on pseudorapidity.

\section{Prediction of the ionization pulse shape}
\label{sect:prediction}

The LAr calorimeters are equipped with a calibration system to inject
an exponential pulse of precisely known amplitude onto intermediate ``mother" 
boards located on the front face (for layer 1) and
back face (for layers 2 and 3) of the calorimeter. The exponential decay 
time of these calibration signals
has been trimmed to mimic the triangular ionization pulse shape as
closely as possible. Since the readout path of the calibration signals is
identical to that of the ionization pulses, the gain and pulse response of 
the electronics can be measured with the calibration system
over the full range of signal amplitudes and time delays. The exponential
calibration pulse properties are analytically modeled via two
parameters $\tau_{cali}$ (inverse of the exponential slope) and $f_{step}$ 
(relative amplitude of a voltage step coming together 
with the main exponential signal).

The main ingredient needed for accurate energy and time reconstruction in the
LAr EM calorimeter is the precise knowledge of the ionization signal shape
in each readout cell, from which the optimal filtering coefficients~\cite{cleland}
are computed. This knowledge of the ionization pulse shape is also 
necessary to accurately equalize the response across cells to account for its difference 
in shape and amplitude with respect to the calibration pulse. The difference between 
the two pulses is due to the slightly different shape of the induced current 
(triangle versus exponential) and the different injection point for the currents 
(electrode versus mother board).

The prediction of the ionization pulse shape relies on the modeling of
each readout cell as a resonant $RLC$ circuit (where $C$ corresponds
to the cell capacitance, $L$ to the inductive path of the ionization
signal and $R$ to the contact resistance between the detector cell and
the readout line) and on the description of the signal propagation
including reflections, amplification and shaping by the
readout electronics.

In the standard ATLAS pulse shape prediction method, Response Transformation
Method (RTM)~\cite{rtm}, calibration pulses are used to determine the description
of the signal propagation and the response of the readout electronics,
as well as the parameters describing the electrical properties of the readout cell, 
($LC$ and $RC$) and the calibration signal ($\tau_{cali}$ and $f_{step}$).

A second method has been developed for the EM barrel, 
First Principles Method (FPM)~\cite{fpm}, 
where the signal propagation and the response of the readout electronics are 
analytically described, and the goodness of the analytical description is 
tuned using the measured calibration pulses.

Both methods need, as an input parameter, the value of the drift time in
each cell, which can be either inferred from the local geometry
of the detector along with the actual LAr temperature and high voltage, or
measured from data pulses as described in this work. Details on the
two methods, which describe the ionization pulse equally well, are given below.

\subsection{RTM Method}
\label{sect:rtm}

The properties of the signal propagation and of the electronic response of the 
readout of the LAr EM calorimeter cells are probed by the
calibration system and can be determined from the measured calibration
pulses. The two underlying assumptions behind the RTM~\cite{rtm} are that:

\begin{itemize}

\item the ionization pulse ($g_{phys}$) can be numerically predicted from the corresponding
      calibration pulse ($g_{cali}$) by means of time domain convolution with two
      simple functions, parameterizing respectively the shape
      difference between the exponential and triangular currents, and
      their different injection points in the detector, see~\cite{rtm}:

\begin{eqnarray}
g_{phys}(t) &=& g_{cali}(t)  \nonumber \\
  &*& {\mathcal{L}}^{-1} 
  \left\{\frac{(1+s\tau_{\rm cali})(s T_{drift}-1+e^{-sT_{drift}})}
  {s T_{drift}(f_{\rm step}+s\tau_{\rm cali})}\right\}  \nonumber \\ 
  &*&{\mathcal{L}}^{-1} 
  \left\{\frac{1}{1+s^2LC+sRC}\right\}  
\label{eq:PulsePred}
\end{eqnarray}

\noindent where $\mathcal{L}^{-1}$ denotes an inverse Laplace transform, 
with $s$ being the variable in the frequency space. 
The first time-domain convolution corrects for the different
signal shapes through the calibration pulse parameters $\tau_{cali}$
and $f_{step}$ and the drift time $T_{drift}$, while the second convolution
accounts for the different injection points on the detector cell,
modeled as a lumped $RLC$ electrical circuit.

\item all parameters ($\tau_{cali}$, $f_{step}$, $LC$, $RC$) used in the convolution functions, 
	apart from the drift time, are directly extracted from the measured
      calibration pulses by numerical analysis~\cite{rtm}.

\end{itemize}

\subsection{FPM Method}
\label{sect:fpm}

  In the FPM method, the signal generation is based on ``first principles" of signal
propagation~\cite{fpm}.  All the calculations are made in the frequency domain, and
when the signal at the output of the final shaping amplifier is obtained, it is transformed to the
time domain by using a fast Fourier transform~\cite{fftw}.

  After generation at the detector level, a signal is propagated along the signal cable, taking into
account its impedance, propagation velocity, and absorption by the skin effect~\cite{fpm}.
A small fraction of this signal is reflected at the signal cable-feedthrough transition, while 
the rest is transmitted. A second reflection takes place at the feedthrough-preamplifier transition.
In this model, the feedthrough is modeled as a single cable section, with its own impedance, 
skin effect absorption constant, propagation velocity and length.
The preamplifier is described by a complex impedance, the real part and the imaginary part
($Re[Z_{PA}]$, $Im[Z_{PA}]$) being both functions of the frequency $\omega$.
The last element of the chain is the $CR-RC^2$ shaping amplifier, described 
by the transfer function:

\begin{equation}
 F_{sh}(\omega)=\omega \cdot \tau_{sh}/(1+(\omega \cdot \tau_{sh})^2)^{3/2}
\end{equation}

\noindent where $\tau_{sh}$ is the $RC$ time constant of this element.
The model accounts for both the directly transmitted signal and the reflections up to the second order
(i.e. two forward-backward reflections and two backward-forward reflections). 

     Parameters are taken from construction (cable lengths, $f_{\rm step}$ and $\tau_{\rm cali}$, 
     which were measured for all calibration boards~\cite{board}), from direct
measurements channel-by-cha\-nnel (resonance frequency $\omega_0 = 1/\sqrt{LC}$ and $R$)~\cite{baffioni},
and from measurements on representative samples 
($Re[Z_{PA}]$, $Im[Z_{PA}]$, propagation velocity and skin effect constants).
The signal cable impedance $Z_S$ and the shaper time constant $\tau_{sh}$ were left as free 
parameters and fitted channel-by-channel on calibration pulses~\cite{fpm}.
The values obtained for $Z_S$ and $\tau_{sh}$ came out close to expectations,
giving confidence in the method which describes calibration pulses to $1 \%$ or better.
The relative timing of all calibration signals was also reproduced with an accuracy of about $1 \rm \ ns$. 

This method was not extended to the EM endcap because not all the necessary parameters have been measured
with the required precision due to a more complex geometry.

\begin{table*} [htbp]
  \centering
  \begin{tabular}{lllll}
    \hline \hline
    & Layer & $S_{max}$ (ADC count) lower limit  & $\sigma_{noise}$ (ADC count) & $F$(MeV/ADC count)\\
    \hline
    \multirow{5}{*}{Barrel} 
    & Presampler & 200 & $8.0$ & $7.0$\\
    & Layer 1 & 500 & $8.0$ & $2.5$\\
    & Layer 2 ($|\eta|\leq 0.8$) & 160 & $5.0$ & $10.0$\\
    & Layer 2 ($|\eta|> 0.8$) & 100 & $3.5$ & $17.0$\\
    & Layer 3 & 160 & $5.0$ & $7.0$\\
    \hline
    \multirow{3}{*}{Endcap} 
    & Layer 1 & 500 & $7.0$ & $3.0$\\
    & Layer 2  & 160 & $4.0$ & $14.0$\\
    & Layer 3 & 160 & $2.0$ & $7.0$\\
    \hline \hline
  \end{tabular} 
  \caption{Cut values for the most energetic sample of the data pulse. 
    The approximate electronic cell noise ($\sigma_{noise}$) 
    averaged over layer and the approximate multiplicative
    conversion factor from ADC counts to MeV ($F$)
are given as well.}
  \label{table:energycut}
\end{table*}

\section{Description of the data}
\label{sect:data}

Cosmic muon runs from the data-taking period of September - November 2008
are used in this analysis, corresponding to a period where the LAr data acquisition 
system transmitted and saved 32 samples of the readout signals. The events of 
interest are those where muons lose a substantial fraction of their energy by 
radiation (the energy lost by $dE/dx$ in layer 2 is in average about $300 \rm \ MeV$~\cite{readiness}). 
These events were triggered using calorimeter trigger towers over 
the full calorimeter depth, of size $\Delta \eta \times \Delta \phi = 0.1 \times 0.1$ 
for $|\eta|<2.5$, $0.2 \times 0.2$ for $2.5 < |\eta| < 3.2$, and up to $0.4 \times 0.4$
for $3.2 < |\eta| < 4.9$. Since the data were collected from
cosmic muons instead of LHC collisions, trigger thresholds were adjusted accordingly.
For technical reasons, only cells which were readout in high gain (LAr readout has three
gains with ratio $\sim 100$/$10$/$1$) are selected for this analysis. This has
a very small impact on the selected sample as the energy deposits are typically 
in the high gain range (energies below $20 \rm \ GeV$).

Despite the small rate of cosmic muons depositing significant electromagnetic energy, 
the number of events recorded during the run period
ensured sufficient statistics for most of the calorimeter
regions, with the exception of the high-$\eta$ region of the endcaps.
The pseudorapidity range in this study is hence restricted to $|\eta| < 2.5$.
 
To minimize distortion of the signal shape, the energy deposited
in a cell is required to be well above its typical noise value. This is
particularly important since the drift time is obtained on an event-by-event basis. 
The quantity $S_{max}$ is defined as the
amplitude of the most energetic sample of the data pulse. The minimal
required values for $S_{max}$ are given in Table~\ref{table:energycut}
for the different layers of the calorimeter; these values correspond to
about $1-2 \rm \ GeV$. The average noise is also quoted, representing
between $1$ and $4 \%$ of the minimal value for $S_{max}$. 
Unless differently stated, all ADC values are pedestal subtracted.
The difference of thresholds between the $|\eta|< 0.8$ and $0.8<|\eta|<1.4$ regions
in layer 2 of the barrel is required by a difference in gain. To correct for
this effect, the normalized variable $S_{max}^{gain}$ is used for the selection, defined as
$S_{max}^{gain}=1.6 \cdot S_{max}$ for $0.8<|\eta|<1.4$,
and $S_{max}^{gain}=S_{max}$ everywhere else. An upper limit of 3900 ADC counts for $S_{max}^{gain}$ plus pedestal is also
required to avoid saturation.

As a small fraction of the ionization pulses are distorted and their
 drift times cannot be determined accurately, a set of cuts has been 
defined to select good quality pulses:

\begin{itemize}

\item The data should have a negative undershoot in the pulse shape. This is ensured 
  by requiring that at least 5 samples after the peak have a negative amplitude.

\item In order to prevent pulses with too short an undershoot
  (as can be the case for signals resulting from crosstalk for instance), 
  a condition requires that the pulse does not contain 
  more than 12 samples around $0 \rm \ ADC$ counts at the end of the pulse.
  This condition cannot be applied to the endcap where such shapes occur 
  due to smaller drift-time values at high pseudorapidity.

\end{itemize}

For a small fraction ($6\%$) of the LAr EM calorimeter the high voltage cannot be safely set to the nominal value.
The cells belonging to these regions are excluded in the following. 
The numbers of pulses per layer after quality cuts are given in Table~\ref{table:numbers}. 

\begin{table} [htbp]
  \centering
  \begin{tabular}{lll}
    \hline \hline
    & Layer & $\#$ pulses  \\
    \hline
    \multirow{4}{*}{Barrel} 
    & Presampler & $20 \rm \ k$  \\
    & Layer 1 &      $43 \rm \ k$ \\
    & Layer 2 &     $331 \rm \ k$ \\
    & Layer 3 &       $79 \rm \ k$ \\
    \hline
    \multirow{3}{*}{Endcap} 
    & Layer 1 &      $13 \rm \ k$   \\
    & Layer 2  &    $45 \rm \ k$   \\
    & Layer 3 &       $18 \rm \ k$   \\
    \hline \hline
  \end{tabular} 
  \caption{Approximate number of cosmic muon induced pulses in each layer after quality cuts.}
  \label{table:numbers}
\end{table}

\section{Extraction of the drift time}
\label{sect:fit}

The 32 data samples $S_{i}$ of each calorimeter cell selected by the criteria 
given in Section~\ref{sect:data} are fitted using the pulse predictions described 
in Section~\ref{sect:prediction}, scaled by an amplitude $A_{max}$ and 
shifted in time by an offset $t_0$:

\begin{eqnarray}
  g_{fit}(t;A_{max},t_0,T_{drift},x) =   A_{max} \cdot \nonumber \\
 g_{phys}(t;f_{nom},T_{drift},x,f_{bend},T_{bend}) \hspace{0.5cm} {\rm for} \ \ t>t_{0}
  \label{eq:TdriftFit}
\end{eqnarray} 

Four parameters are left free in this procedure:
the drift time ($T_{drift}$), 
the associated shift of the electrode estimated as $\delta_{gap}$ $=$ $x \cdot w_{gap}$
which is in fact only sensitive to the absolute value of $x$ when the high voltage is 
the same on both sides of electrodes, 
the global normalization factor $A_{max}$ and the timing adjustment $t_{0}$. 
The optimal set of these four parameters is estimated using the least squares method to minimize
the quantity:
\begin{equation}
  Q^{2}_{0} =  \frac{1}{n - N_p} \sum_{i=1}^{n}
  \frac{ 
    \left(S_i - g_{fit}(t_i;A_{max},t_0,T_{drift},x) \right)^2 
  }
  { \sigma^2_{noise} }
  \label{eq:chi2fit}
\end{equation}
where $n$ is the total number of data samples used in the fit (usually
$n=32$), $N_p$ the number of free parameters ($N_p=4$), and
$\sigma_{noise}$ is given in Table~\ref{table:energycut}. This minimization
is performed using the MINUIT package \cite{MINUIT}.

\begin{figure}[htbp]
\begin{center}
\subfigure[$Q^{2}_{0}$ versus $S_{max}^{gain}$]{
\label{F:chi2}
\includegraphics[width=0.8\columnwidth]{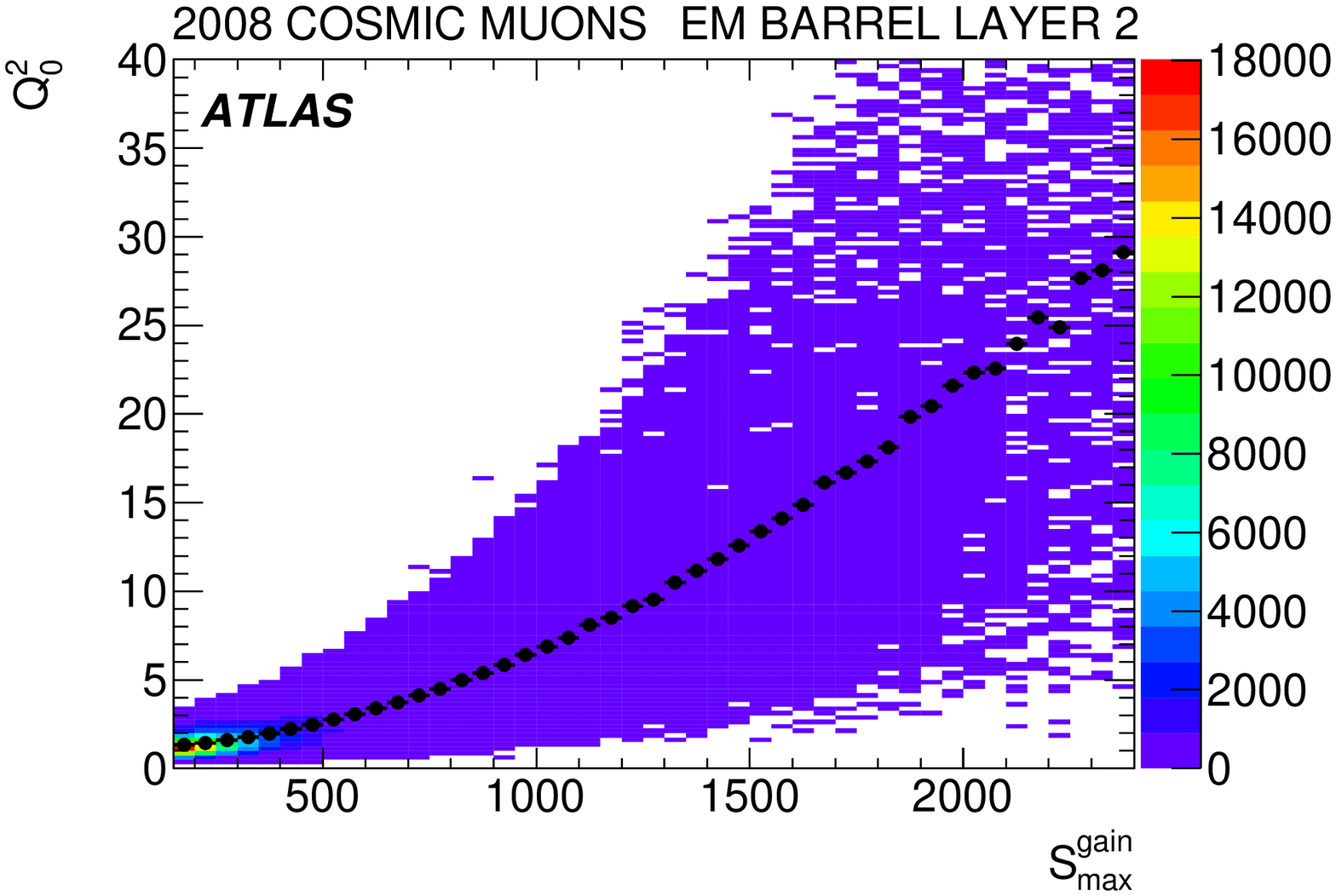}
}
\subfigure[$Q^{2}$ versus $S_{max}$]{
\label{F:chi2star}
\includegraphics[width=0.8\columnwidth]{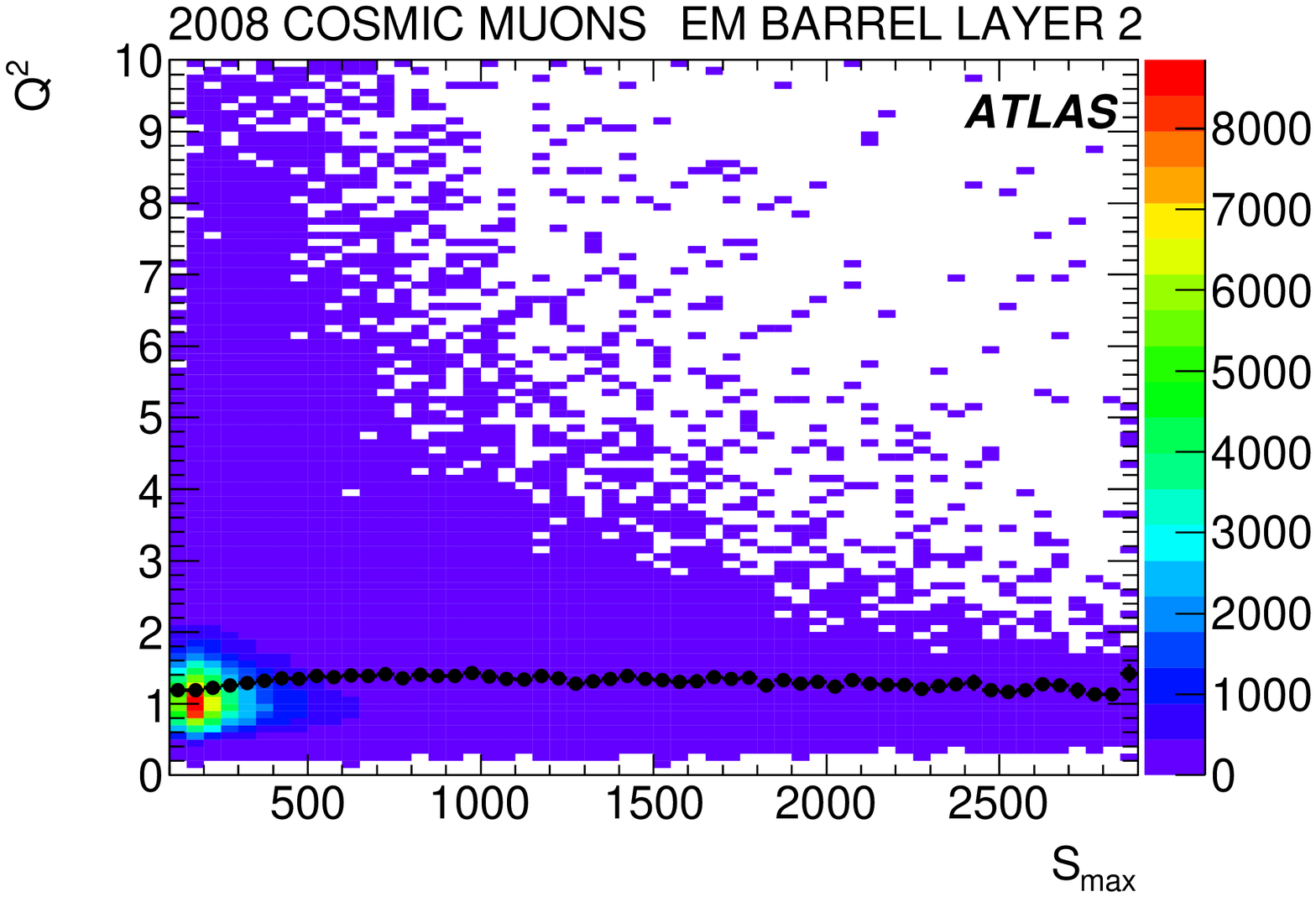}
}
\caption{(a) $Q^{2}_{0}$ versus $S_{max}^{gain}$ and
(b) $Q^{2}$ versus $S_{max}$ in layer 2 of the barrel.
The black points correspond to the mean value.
\label{F:chi2_distri}}
\end{center}
\end{figure}

\begin{table*} [htbp]
  \centering
  \begin{tabular}{llll}
    \hline \hline
     Layer & $k_{\rm FPM}$ in barrel & $k_{\rm RTM}$ in barrel & $k_{\rm RTM}$ in endcap \\
    \hline
     Presampler & 0.9\% &    &   \\
     Layer 1      & 1.1\% & 0.8\% & 0.9\%  \\
     Layer 2     & 0.8\% & 1.0\% & 1.4\%  \\
     Layer 3      & 0.75\% & 1.0\% & 1.3\%  \\
    \hline \hline
  \end{tabular}
  \caption{$k$ values for the different methods in the different regions of the EM calorimeter.}
  \label{table:kvalues}
\end{table*}

Figure~\ref{F:chi2} presents the variation of $Q_{0}^2$ with
$S_{max}^{gain}$ for layer 2 of the barrel. An increase of the $Q_{0}^2$
value is observed when $S_{max}^{gain}$ is larger. The same behavior is observed in 
the other calorimeter layers, as expected. In order to be able to apply 
a global selection to the fit quality independently of the data amplitude, 
a ``normalized'' $Q_{0}^2$, called $Q^{2}$, has been used: 

  \begin{equation}
  Q^{2} = \frac{1}{n - N_p} \sum_{i=1}^{n} 
  \frac{ 
    \left(S_i - g_{fit}(t_i;A_{max},t_0,T_{drift},x) \right)^2 
  }
  { \sigma^2_{noise} + (k S_{max})^2 }
\label{eq:schi2}
\end{equation}

\noindent where $k$
is chosen such that $Q^{2}$ is independent of
$S_{max}$, as represented in Figure~\ref{F:chi2star}.
The denominator in Equation~\ref{eq:schi2} is the quadratic sum of the noise 
and of the relative inaccuracy of the
predicted shape. It represents the numerator uncertainty.
The values of $k$ are given in Table~\ref{table:kvalues}
for the different layers of barrel (two methods) and endcap.

For the measurement of the drift time, the last data samples 
corresponding to the end of the pulse are very important. It was noticed 
that for a small fraction of pulses ($\sim 0.6 \ \%$ for the layer 2) 
the fit converges successfully but the 
predicted pulse does not succeed in describing the rise at the end of the pulse. 
This implies an incorrect estimate of the drift time. 
To specifically quantify the quality of the fit at the end
of the pulse, the variable $\Delta_{\rm last7}$ has been defined, based only on the last 7
samples: 

  \begin{equation}
    \Delta_{\rm last 7} = \sum_{i=26}^{32} 
    \frac{S_i - g_{fit}(t_i;A_{max},t_0,T_{drift},x)}{S_{max}}
    \label{eq:Dlast7}
  \end{equation}
  
\noindent Large values of $|\Delta_{\rm last7}|$ single out pulses with erroneous fitted drift times.
This effect is also observed with a toy simulation, and therefore seems to be an intrinsic 
feature of the fitted function, with a large peak followed by a flat tail. 

\begin{figure*}
\begin{center}
\subfigure[Layer 2 of the barrel]{
\label{F:param_barrel}
\includegraphics[width=0.8\columnwidth]{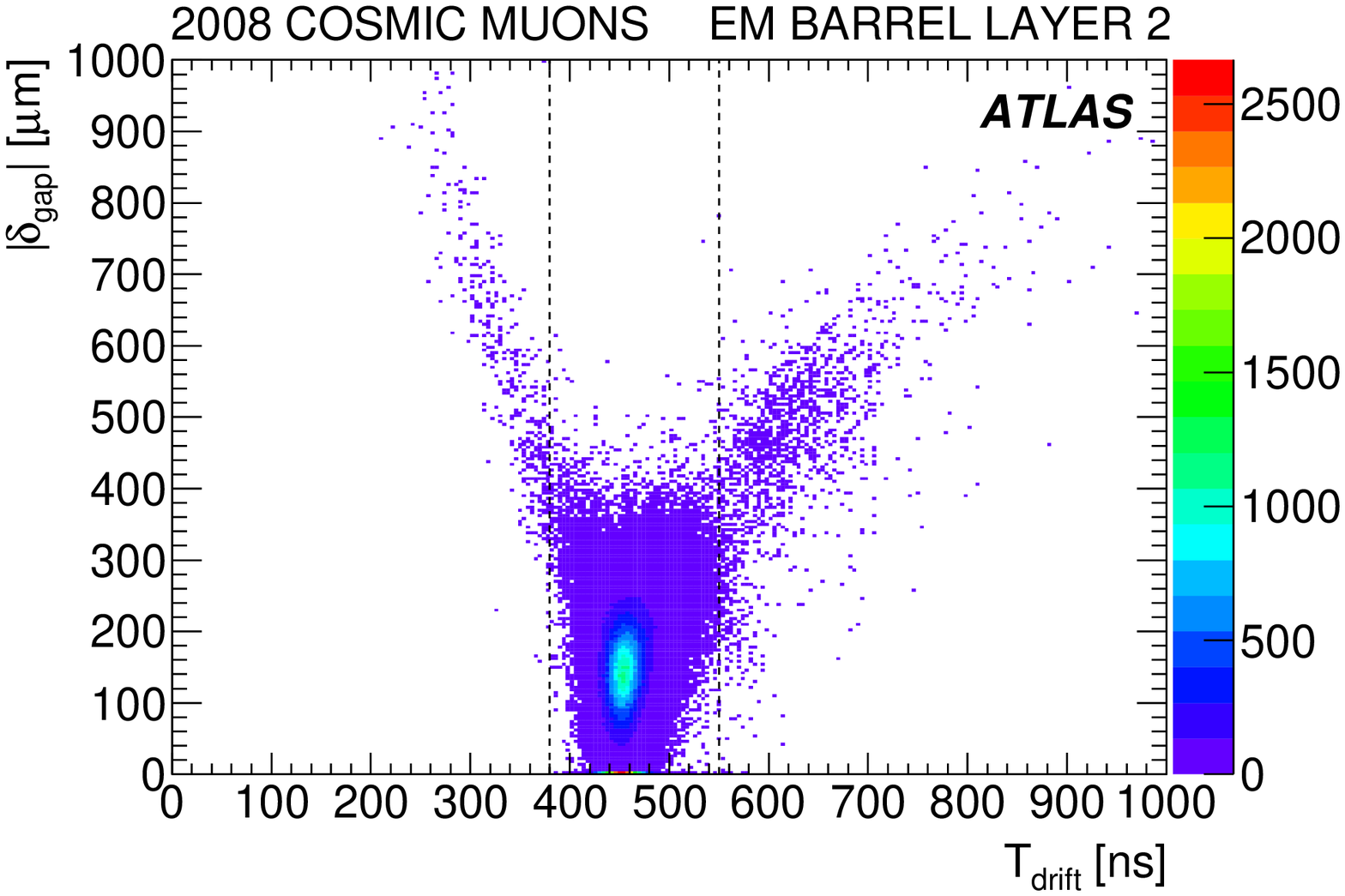}
}
\subfigure[Layer 2 of the endcap]{
\label{F:param_endcap}
\includegraphics[width=0.8\columnwidth]{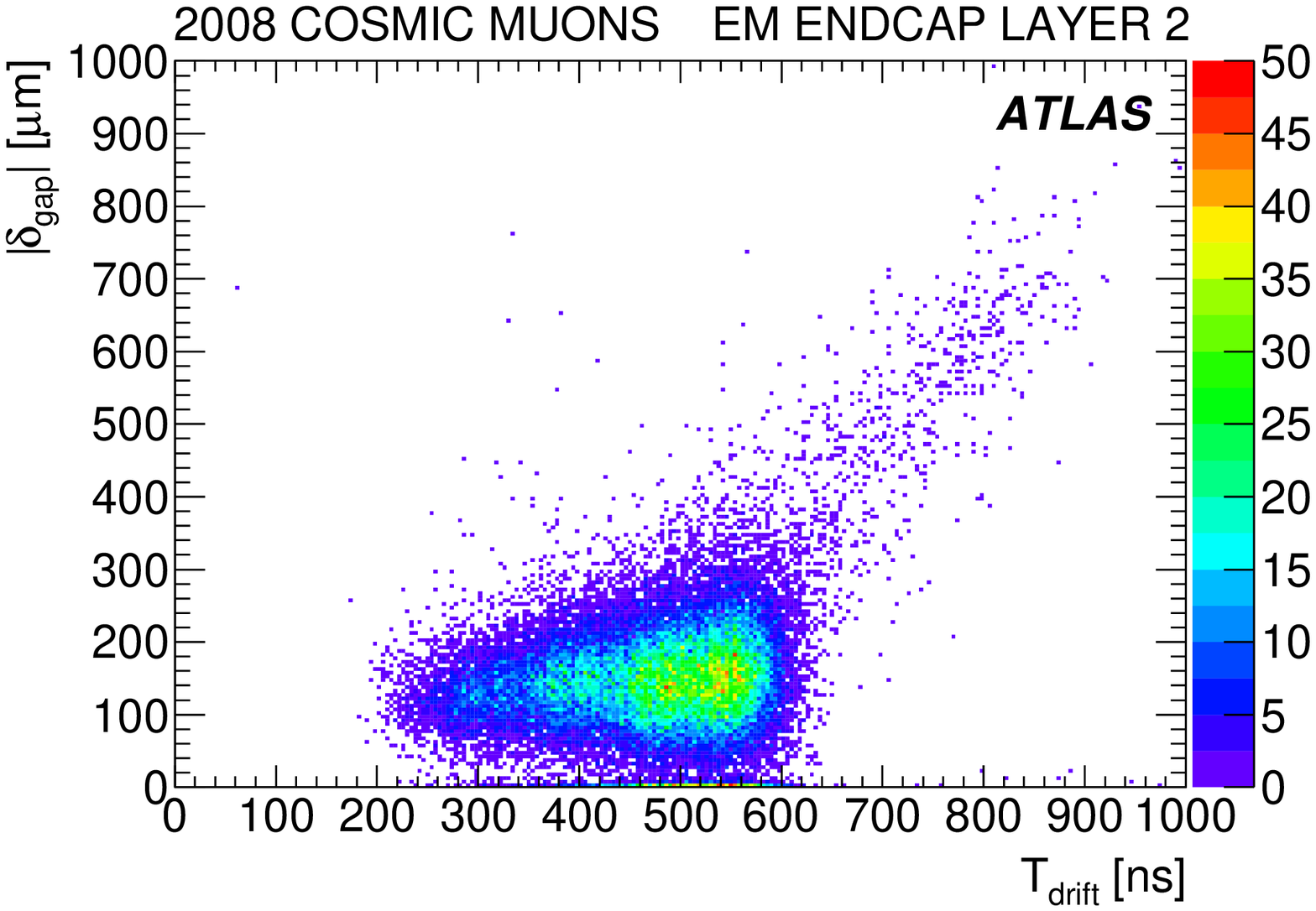}
}
\caption{Absolute value of the shift parameter as a function of the drift time in the barrel (a) and
in the endcap (b), for layer 2.
\label{F:params}}
\end{center}
\end{figure*}

\begin{figure*}
\begin{center}
\subfigure[Layer 2 of the barrel]{
\label{F:shift0_barrel}
\includegraphics[width=0.8\columnwidth]{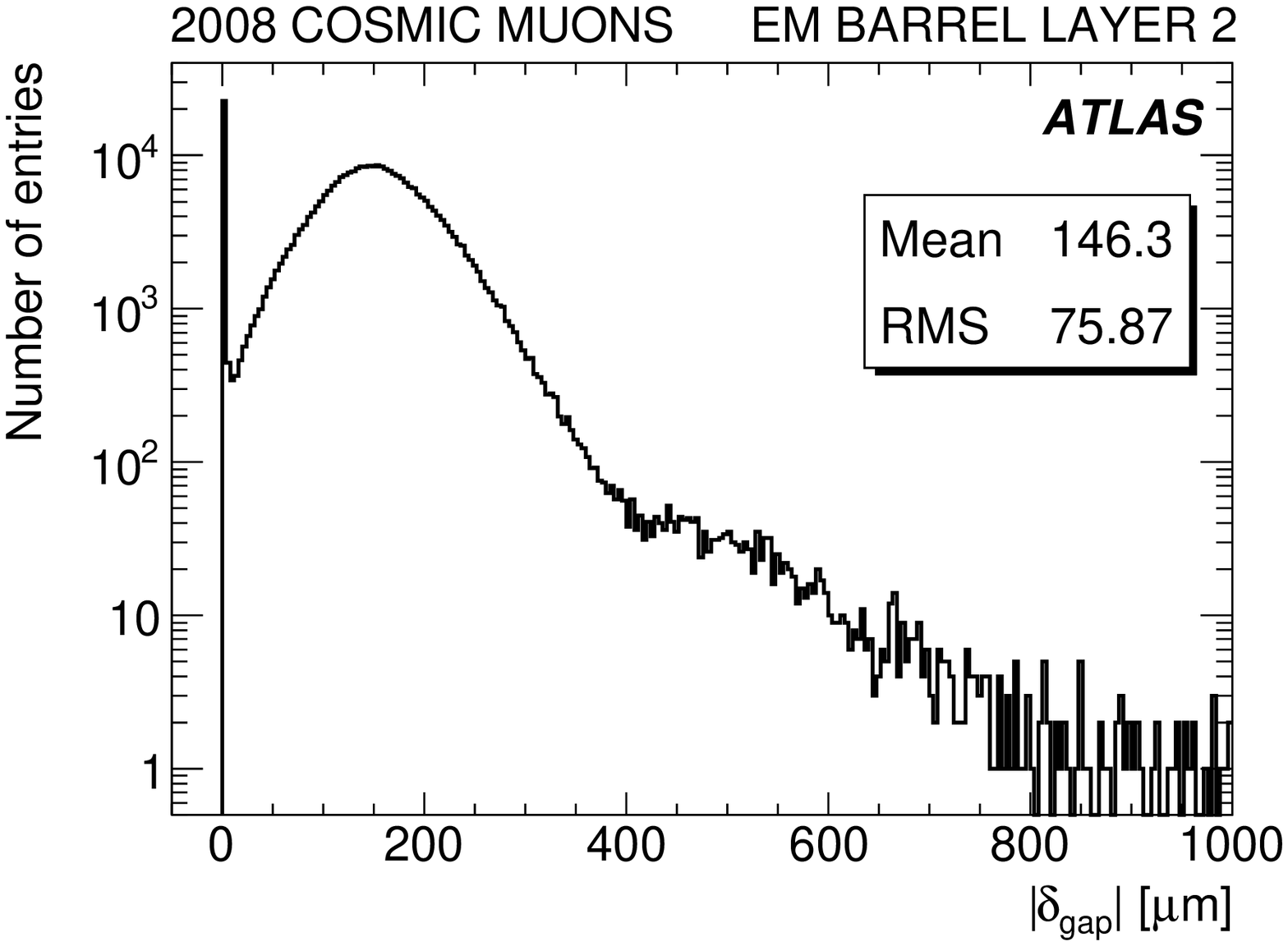}
}
\subfigure[Layer 2 of the endcap]{
\label{F:shift0_endcap}
\includegraphics[width=0.8\columnwidth]{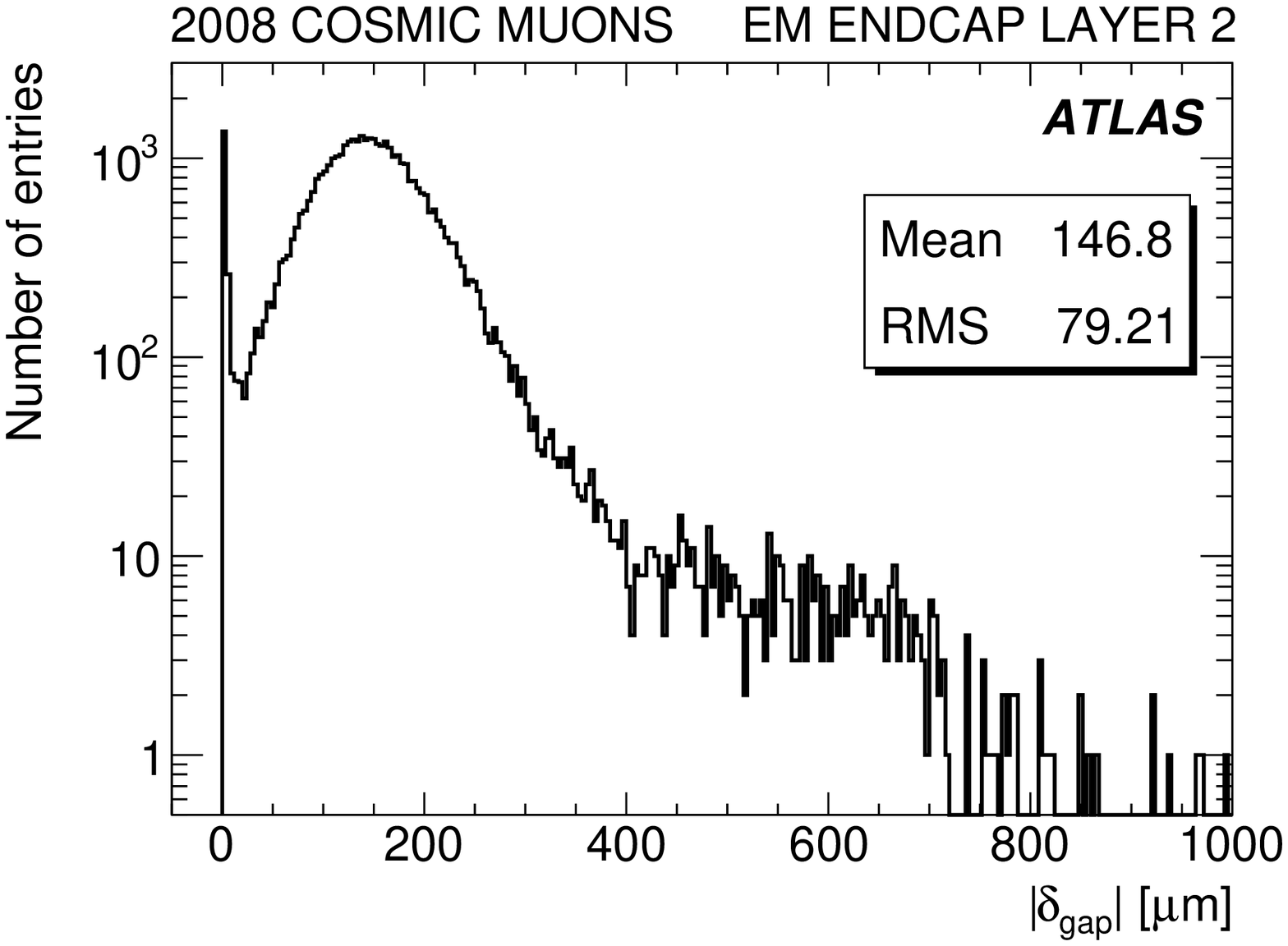}
}
\caption{Distribution of the absolute value of the shift parameter in layer 2 of the barrel (a) and endcap (b).
\label{F:shift0}}
\end{center}
\end{figure*} 

To remove events for which the end of the pulse is badly described by the model, a cleaning selection requiring
$|\Delta_{\rm last7}|<0.15$ and $Q^{2} < 2.5$ (3) in the barrel (endcap) is imposed.

An additional set of cuts on the maximum relative residual over all samples is applied for presampler cells,
where pick-up of oscillatory signals was in a few places observed ($3 \ \%$ of the pulses):

\begin{itemize}
  \item $|S_i-g_{fit}(t_i)|_{max}/S_{max} < 10\%$,
  \item if the residual is small ($|S_i-g_{fit}(t_i)|_{max} < 20 \rm \ ADC$ counts), the cut is relaxed to
$|S_i-g_{fit}(t_i)|_{max}/S_{max} < 20\%$.
\end{itemize}

 After these selections, the fit parameters are examined in more detail.
Figure~\ref{F:params} presents the distribution of the absolute value of the
shift parameter, $\delta_{gap} = x w_{gap}$, as a function of the drift time. 

The region in Figure~\ref{F:param_barrel} with a drift time $T_{drift}$
comprised between $380$ and $550 \rm \ ns$ corresponds to the expected 
range for the drift time in the barrel given the resolution of the measurement.
The low drift time region $T_{drift} <380 \rm \ ns$ of
Figure~\ref{F:param_barrel} ($0.05 \%$ of the pulses)  is dominated by low-amplitude
pulses distributed evenly in the calorimeter. A closer examination shows that in about $80\%$ 
of the cases for the layer 2 barrel, signals in excess of $S_{max}=1500 \rm \ ADC$ 
counts or cells sampled at medium gain are found as first neighbors which corroborates a crosstalk hypothesis.

In the region $T_{drift}>550\rm \ ns$ of Figure~\ref{F:param_barrel} ($0.25 \%$ of the pulses), some pulses are still
significantly  negative, more than $700\rm \ ns$ after the time of signal
maximum. A possible explanation is that the energy deposit originates from a photon
emitted along a bent section, thus having an abnormally enhanced $f_{bend}$
contribution.  Unfortunately the runs taken with $32$ samples
do not contain information from the inner tracker which would have
allowed this hypothesis to be validated by a projectivity study. Aside from these
extremely large drift time pulses, there is a larger class of
pulses which are only somewhat longer than normal.
They are distributed along specific $\eta$ and $\phi$ directions:
in the transition regions at $|\eta|= 0.8$ and between the two half-barrels at $\eta =0$ (see Section~\ref{sect:bmid})
where a slight dilution or leakage of the
electric field lines yields a larger drift time
(this is also observed in layer 1 of the barrel);
in the intermodular regions in $\phi$ in the upper part of the detector,
where mechanical assembly tolerances allow for a slightly increased gap 
at the interface between modules due to gravity effects (this is not seen in 
barrel layers 1 and 3 which are much closer to the mechanical fixed points). 

In the endcap, the cloud of points corresponding to the
expected $T_{drift}$ is broader than in the barrel, as can be seen in Figure~\ref{F:param_endcap}:
it ranges from 300 to 600 ns as a consequence of the gap size variation with $\eta$
of the endcap design. The fact that the dispersion of $|\delta_{gap}|$ is larger at
higher values of $T_{drift}$ is explained as a consequence of the larger
gap size: the larger the gap width, the larger the displacement of the electrode can be.
A few events ($0.9 \%$ of the pulses) are observed at very high values of both $T_{drift}$ and
$|\delta_{gap}|$. They are located at low $|\eta|$
where the drift time is very large by construction (see Figure~\ref{F:mc_endcap_td}). 
Their pulse shape cannot be completely readout using 32 samples, and in particular
the rise following the undershoot is partially absent, which leads
to unphysical values of the shift above $400 \mum$. 

A distinctive aspect of the fit, which is clearly visible in
Figure~\ref{F:shift0},  is that it yields a peak at $|\delta_{gap}|=0$. 
This is mainly explained by noise fluctuations. The superposition of two 
triangles of ionization current with unequal length due to an electrode 
shift (see Figure~\ref{F:current}) can only lead to a softening 
of the rise at the end of the pulse, compared to the single-triangle case. 
If, due to noise, the rise is steeper than for a single-triangular shape, 
the fit forces $\delta_{gap}$ to $0$. In order to improve the statistical 
significance of high-amplitude signals and minimize the impact of noise 
fluctuations, it has been decided to weight the events by $(S_{max}^{gain})^2$. 
The results in the following sections of this note are produced with this 
weighting factor.

\section{Results in the calorimeter barrel}
\label{sect:barrel_result}

Two parallel analyses have been performed for this part of the calorimeter
using the two pulse shape prediction methods described in Section~\ref{sect:prediction}. The analyses
agree at the level of $0.3 \%$, which provides a good check of the robustness of these results. In this section 
the measurement of the drift time is presented, along with its implications for 
the calorimeter response uniformity and an estimation of the electrode shift.

\subsection{Drift time measurement in pseudorapidity and azimuthal angle}

Results in layer 2 are presented first because the statistical uncertainties are lower
 in this layer (see Table~\ref{table:numbers}).
More refined comparisons between the two methods are then possible. The following subsection reports on the results in the 
other layers. The presampler is discussed separately due to its different structure.

\subsubsection{Layer 2 of the barrel}
\label{sect:bmid}

\begin{figure}[htbp]
  \begin{center}
    \includegraphics[width=0.80\columnwidth]{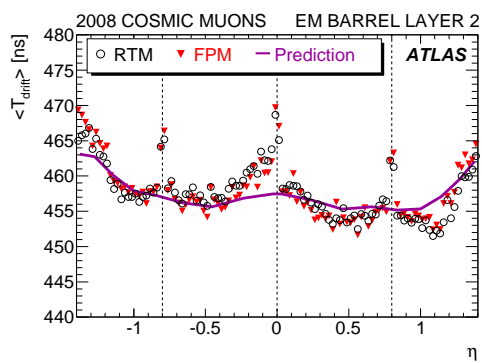}
  \end{center}
  \caption{Drift time as a function of $\eta$ in layer 2 of the barrel: using the RTM method (open dots), the FPM method
	(red triangles) and the prediction described in the text (purple line).
  \label{F:td_barmi_eta}}
\end{figure}

Figure~\ref{F:td_barmi_eta} presents the  drift time $T_{drift}$  extracted from the fit 
as a function of $\eta$.
The results of the two methods differ by $0.1 \rm \ ns$ on average with an RMS of $1.3 \rm \ ns$. 
The full purple line illustrates the prediction from absorber
thickness measurements made during the calorimeter construction~\cite{construction}.
This prediction is based on the fact that the mechanical structure of the calorimeter ensures that the pitch 
(with nominal values shown in parentheses) is constant to within about $5 \rm \mum$:

	\begin{eqnarray}
	{\rm Absorber}\,(2.2\mm) + w_{gap} \,(2.09\mm) \nonumber \\
	+ {\rm Electrode} \,(0.280\mm)  + w_{gap} \,(2.09\mm) = \nonumber \\
	6.66\mm = (2 \pi/1024) \cdot R_i \cos \theta_i
	\end{eqnarray}

\noindent where $R_i$ and $\theta_i$ are the average radius and the local angle of the $1024$
accordion-shaped absorbers with respect to the radial direction.
So if the thickness of the absorber varies with $\eta$, the gap
will also vary in the opposite direction.
As the drift time $T_{drift}$ is directly related to the gap by:
	\begin{equation}
	T_{drift} = T_{D0} (w_{gap}/ w_{gap0})^{1+\alpha}
	\label{eq:gap2td}
	\end{equation}
a prediction can be derived for the drift time from the
variations around the nominal gap
size ($w_{gap0} = 2.09 \rm \ mm$) associated with $T_{D0}=\langle T_{drift} \rangle=457.9 \rm \ ns$.

\begin{figure}[htbp]
  \begin{center}
    \includegraphics[width=0.80\columnwidth]{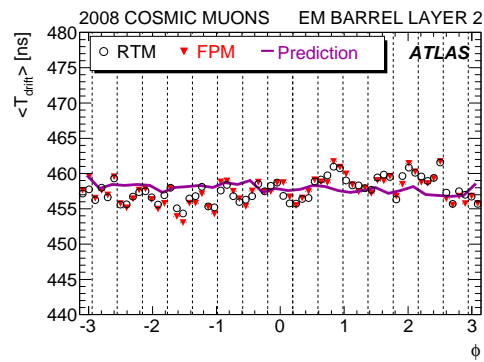}
  \end{center}
  \caption{Drift time as a function of $\phi$ in layer 2 of the barrel: using the RTM method (open dots), the FPM method
	(red triangles) and the prediction described in the text (purple line).
  \label{F:td_barmi_phi}}
\end{figure}

      The agreement between the prediction coming from precision mechanical probe measurements 
of the absorber thickness and the data is rather good,
except in the transition regions around
$\eta=0$, $\pm 0.8$ and $- 1.4$, where the lower field induces a
larger $T_{drift}$.  To quantify the
agreement between the drift time measurements from the fit and the estimate from the 
measurement of the absorbers, the RMS of the difference between the data points and the prediction
is computed. This yields a value of $2.9 \rm \ ns$, as compared to an RMS deviation with respect to a  
constant value of $3.7 \rm \ ns$, excluding the data points around the transition region in each case.
Comparing bin by bin the drift times obtained (Figure~\ref{F:td_barmi_eta}) for the negative
and positive values of $\eta$, one gets a distribution with a mean of $3.4 \pm 0.2 \rm \ ns$
and RMS of $1.7 \rm \ ns$. The predicted value is $1.5 \pm 0.2 \rm \ ns$.

The $T_{drift}$ distribution as a function of $\phi$ is presented in
Figure~\ref{F:td_barmi_phi}, for both methods. There is a small difference
between the $\phi<0$ ($(456.8 \pm 0.3) \rm \ ns$) and $\phi>0$ 
($(458.3 \pm 0.3) \rm \ ns$) regions: a $(0.3 \pm 0.1) \%$ relative effect
consistent with sagging and pear shape deformation of the calorimeter. 
No significant variations are observed in the absorber thickness measurements.
The distribution of the results is also rather uniform when looking at the two half-barrels separately.
The RMS of the $\phi$ distribution is smaller ($1.8 \rm \ ns$) when the two half-barrels are combined, 
than for the $\eta<0$ ($2.8 \rm \ ns$) and $\eta>0$ ($3.1 \rm \ ns$) half-barrels separately. This may be due 
to the existence of small $\phi$ modulations with opposite phases in the two half-barrels that appear to be more
visible in layer 3 (see Figure~\ref{F:td_barback}).

\subsubsection{Other layers of the barrel}
\label{sect:bother}
   
\begin{figure}[htbp]
  \begin{center}
    \includegraphics[width=0.80\columnwidth]{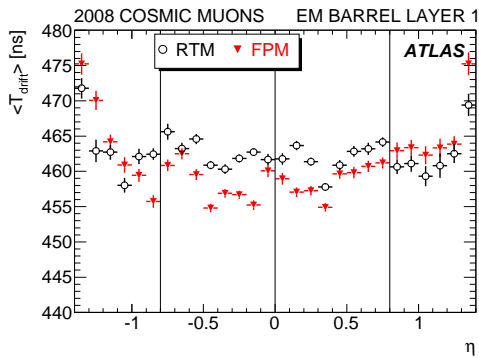}
  \end{center}
  \caption{Drift time as a function of $\eta$ in layer 1 of the barrel: 
  using the RTM method (open dots) and the FPM method (red triangles).
  \label{F:td_barfront}}
\end{figure}

The distribution of $T_{drift}$ as a function
of $\eta$ for layer 1 is displayed in Figure~\ref{F:td_barfront}. 
The results of the two methods differ by $1.3 \rm \ ns$ on average, 
with an RMS of $4 \rm \ ns$, and at some points by
up to $7 \rm \ ns$. The front layer is
particularly intricate because of the large relative variations of the cell
depths which present a discontinuity at $|\eta|=0.8$, inducing a
corresponding variation of the cell capacitance and bent-to-flat 
ratio. Given that the two methods differ in their
estimation of the cell capacitance, such a difference is not unexpected.

\begin{figure}[htbp]
  \begin{center}
    \includegraphics[width=0.80\columnwidth]{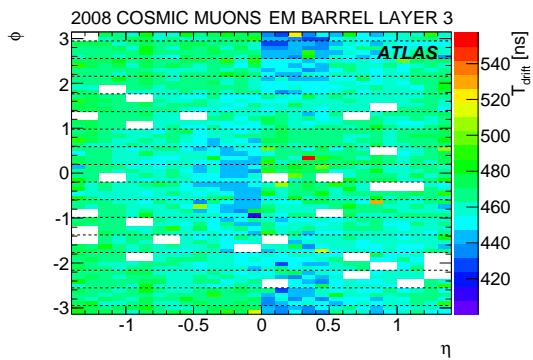}
  \end{center}
  \caption{2D map of $T_{drift}$ in ($\eta$,$\phi$) for layer 3. 
  The empty bins correspond to sectors with non nominal HV.
  \label{F:td_barback}}
\end{figure}

   In Figure~\ref{F:td_barback}, a
drift time  modulation with $\phi$ is clearly visible for $|\eta|<0.5$ 
(and equally present in both methods) in both half
barrels of layer 3. While the source of the modulation is so far unexplained,
the fact that the modulations in the two half-barrels are opposite in phase
is expected, since one of the half-barrels was rotated 
by $180$ degrees about the vertical direction for final integration.

\subsubsection{Presampler}

\begin{figure}[htbp]
  \begin{center}
    \includegraphics[width=0.80\columnwidth]{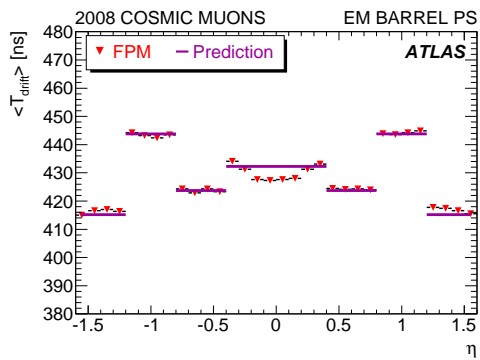}
  \end{center}
  \caption{Drift time as a function of $\eta$ in the presampler barrel
  using the FPM method (red triangles).
The full purple line represents the prediction normalized to the region
$0.8<|\eta|<1.2$, using Equation~\ref{eq:gap2td}
and the gap values given in Table~\ref{tab:presampler}. 
The empty bins correspond to sectors with non nominal HV.
  \label{F:presampler}}
\end{figure}

      The presampler is constructed differently from the other layers of the
calorimeter. It is made of narrow flat electrodes.
The size of the gaps is slightly smaller than elsewhere,  leading to
values of $T_{drift}$ lower than in the rest of the calorimeter. In addition,
this gap varies with $\eta$; the values for the 4 regions are given in
Table~\ref{tab:presampler}.  The effect on the fitted drift time can
be immediately seen in Figure~\ref{F:presampler}.  The prediction
superimposed on the measured distribution is normalized to the region
$0.8<|\eta|<1.2$. Good agreement within $1 \%$ is observed between the
measured and expected drift times as a function of $\eta$. As there
are no bent sections in the presampler, the pulse description is
simpler than in the case of the other layers. While the variations in
$\eta$ are large, the $\phi$ dependence of the drift time is
negligible.

\begin{table}[htbp]
  \centering
    \begin{tabular}{ll}
\hline \hline
$\eta$ region & $w_{gap}$ (in $\rm mm$)\\
\hline
$|\eta|<$ 0.4    & 1.966 \\
0.4 $ \leq |\eta|<$ 0.8  & 1.936 \\
0.8 $ \leq |\eta|<$ 1.2  & 2.006 \\
1.2 $ \leq |\eta|$  & 1.906 \\
\hline \hline
\end{tabular}
\caption{Gap values in presampler.
\label{tab:presampler}}
\end{table}

\subsection{Response uniformity}
\label{S:uni_EMB}
The reconstructed value of the energy deposited in the calorimeter by an electron or photon
should be independent of the position of its impact on the calorimeter. 
The non-uniformity coming from local variations of the response due
to gap fluctuations can be determined using the drift time measurements.
This study is done only for layer 2, which is the main contributor
to the energy response of the detector as it collects  $\sim 70\%$ of
the total electromagnetic signal in the calorimeter.

\begin{figure}[htbp]
  \begin{center}
    \includegraphics[width=0.80\columnwidth]{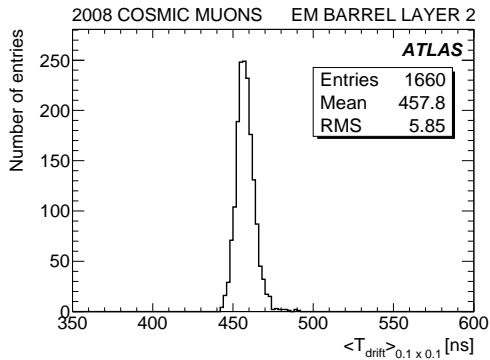}
  \end{center}
  \caption{Drift time uniformity between groups of $4 \times 4$ cells 
  ($\Delta \eta \times \Delta \phi = 0.1 \times 0.1$) for barrel layer 2.
  \label{F:uniformity_middle}}
\end{figure}

Figure~\ref{F:uniformity_middle} shows the distribution of the
drift time averaged over groups of $4 \times 4$ cells corresponding to
an area of $0.1 \times 0.1$ in $\Delta \eta \times \Delta \phi$ plane.
This area represent a typical transverse size of a single particle shower.
The average of the statistical uncertainties on $T_{drift}$ obtained for
 pulses within the various $4 \times 4$ groups is $1.25 \rm \ ns$, well below   
 the dispersion of the determined $T_{drift}$ values of the groups (the RMS
 is $5.85 \rm \ ns$). From the measurement of drift times, the systematic
 dispersion of the gaps can be estimated and its impact on the calorimeter 
 energy response can be assessed.

The drift time uniformity, corresponding to the ratio of the RMS and
the mean value of the local $T_{drift}$ distribution amounts 
to $(5.85 \pm 0.14)/457.8 =(1.28\pm 0.03) \%$.
From the relation between the drift time and the drift 
velocity (Equation~\ref{eq:pure_vdrift}), the latter being proportional to the energy response, together with 
Equation~\ref{eq:gap2td}, it follows, that the drift time uniformity leads to a
dispersion of the response due to the gap variations of 
$(1.28 \pm 0.03) \ \% \cdot (\alpha / (1+\alpha)) = (0.29 \pm 0.01) \%$. Excluding transition 
regions in $\eta$ and in $\phi$, the gap variations amount to $5.7/457.4 = 1.25\%$ and the impact
on the response is $0.28\%$. Taking into account small variations observed in the result
when changing the weighting, the fit strategy (see Section~\ref{sec:systematics})
or the pulse reconstruction method, a systematic error of $0.03 \%$ is obtained.
The uncertainty on $\alpha$ (see Section~\ref{sect:ionization}), treated here as
external parameter, contributes with a systematic uncertainty of $^{+0.04}_{-0.02}$.
Grouping all errors together in quadrature gives as the final result: $ (0.29^{+0.05}_{-0.04}) \ \%$.

\subsection{Electrode shift}

As presented in Section~\ref{sect:ionization}, there is some freedom for the electrodes
to be displaced with respect to their nominal positions equidistant between two neighboring absorbers.
This displacement is expected to be less than $400\rm \mum$ except perhaps in the transition regions between modules.

The electrode shift is left as a free parameter in the fit to the data, which yields
one value per calorimeter cell. Only the average of the absolute value of the displacement 
can be observed.  Since a cell consists of several electrodes, an effective value 
is obtained which is a combination of the individual movements of each electrode 
within a cell. 

\begin{figure}[htbp]
  \begin{center}
    \includegraphics[width=0.80\columnwidth]{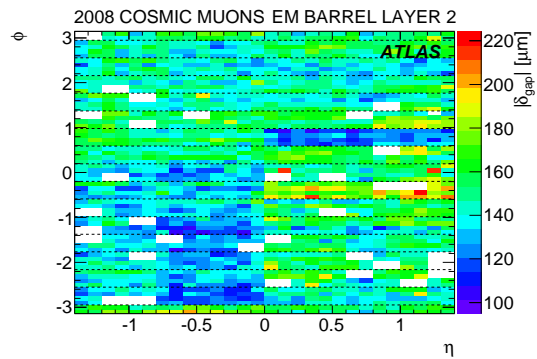}
  \end{center}
  \caption{($\eta$,$\phi$) map in which
$|\delta_{gap}|$ is plotted per bin of $0.1 \times 0.1$.
  \label{F:sh_etaphi_middle}}
\end{figure}

The local average value for the shift parameter 
per bin of $0.1 \times 0.1$ is shown in Figure~\ref{F:sh_etaphi_middle} for layer 2. It indicates that
the bottom half (negative $\phi$) of the negative-$\eta$ half-barrel
has  shift parameter values somewhat lower than average.  Similarly
the module azimuthally located between $4\pi/16$ and $5\pi/16$ in the $\eta>0$
half-barrel presents lower shift values. These variations given 
their distribution throughout the detector, are likely to be due to mechanical
construction issues.

The shift parameter also covers local variations of the ``double-gap" within a cell,
for example, by the slight opening of gaps along an accordion fold. This latter
variation is in general much smaller than the off-centering of electrodes between
absorbers.

Smaller values of the shift parameter are expected for the presampler
compared to the accordion layers, due to mechanical
constraints on the electrodes which are individually glued in between
two precision structural frames \cite{construction}. The mean value of the
shift in the presampler is found to be $\langle|\delta_{gap}|\rangle = 66.5 \rm  \mum$, 
as compared to $146 \rm  \mum$ in the accordion section. 

\section{Results in the calorimeter endcap}
\label{sect:endcap}

As was done for the barrel, the endcap results are grouped in three different 
parts: drift time measurements, calorimeter response uniformity 
and electrode shift determination.

\subsection{Drift time measurement in pseudorapidity and azimuthal angle}

\begin{figure}[htbp]
\begin{center}
\subfigure[Layer 1]{
\label{F:endcap_fr}
\includegraphics[width=0.8\columnwidth]{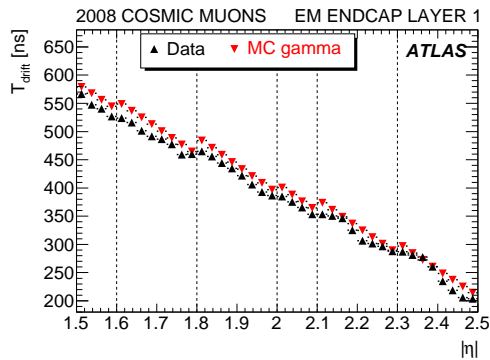}
}
\subfigure[Layer 2]{
\label{F:endcap_mi}
\includegraphics[width=0.8\columnwidth]{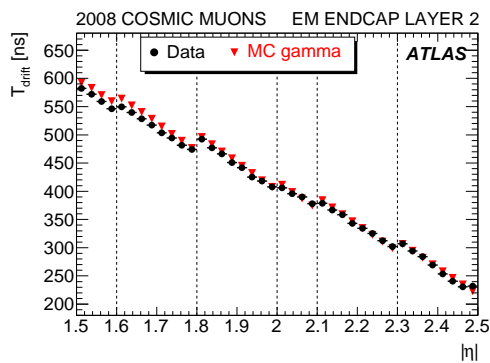}
}
\subfigure[Layer 3]{
\label{F:endcap_back}
\includegraphics[width=0.8\columnwidth]{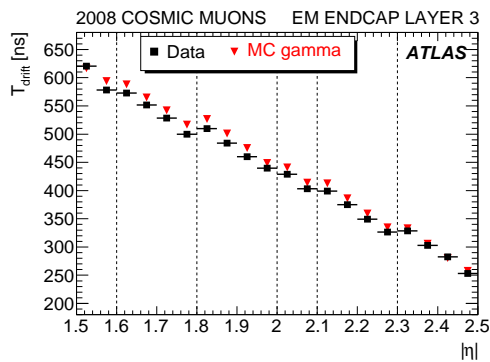}
}
\caption{Drift time versus pseudorapidity for layer 1 (a), layer 2 (b), and 
layer 3 (c) cells  of the endcap.
Black points are the data and red triangles Monte Carlo predictions for photons.
The vertical dashed lines show the boundaries between different high voltage regions. }
\label{F:TdvsetaS1S2S3EMEC}
\end{center}
\end{figure}

The drift time $T_{drift}$ averaged over $\phi$ is studied as
a function of $\eta$ for each of the three layers of the endcap
(see Figure~\ref{F:TdvsetaS1S2S3EMEC}). The two endcaps, A ($\eta>0$) and C ($\eta<0$), are 
combined in the figure. 
A general decrease of $T_{drift}$ with increasing pseudorapidity
is observed, as expected from the corresponding reduction of the design gap size.
Fewer fluctuations are observed in layer 2,
which offers a larger cross section to cosmic muon-induced electromagnetic showers.
In all layers regular steps are
observed, corresponding to the locations of the boundaries between 
 high voltage regions.

The data are compared to the Monte Carlo calculation described in
Section~\ref{sect:ionization}. Good agreement is observed 
at high $\eta$, however the Monte Carlo is slightly above the data at low values 
of $\eta$ ($\sim 1-3 \ \%$), 
which is a more difficult region to simulate.

\begin{figure}[htbp]
  \centering
  \includegraphics[width=0.8\columnwidth]{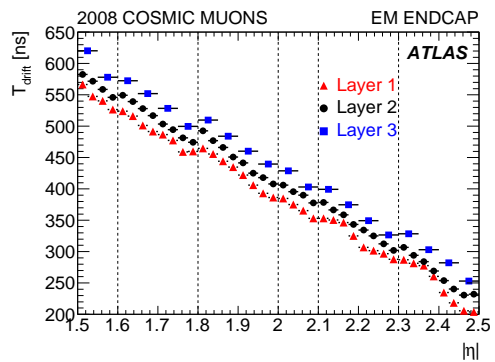}
  \caption{Drift time versus pseudorapidity for the three layers
    of the endcap: layer 1 (red triangles), layer 2 (black dots), layer 3 (blue squares).
    The vertical dashed lines show the boundaries between different high voltage regions.}
  \label{F:Tdvseta3layers}
\end{figure}

In Figure~\ref{F:Tdvseta3layers}, for a comparison, the data points from the three 
distributions of Figure~\ref{F:TdvsetaS1S2S3EMEC} are super-imposed on the same plot.  
An increase of the drift time with the cell gap size
at fixed $\eta$  is clearly observed, with $T_{drift}$ being smallest
for layer 1 and highest for layer 3 (see Section~\ref{sect:ionization} 
and Figure~\ref{F:mc_endcap_td}). The drift
time for layer 2 lies half way between layers 1 and 3
in contrast to the Monte Carlo simulation
(Figure~\ref{F:mc_endcap_td}) where the values for layer 2 are
closer to the values of the layer 1. This difference reflects the fact 
that cosmic muons are randomly distributed within the depth of layer 2, 
while the photons of the simulation develop there
shower closer to layer 1.

\begin{figure*}[htbp]
\centering

\subfigure[layer 2 of endcap A]{
\label{F:tdriftvsphi_EMECA}
\includegraphics[width=0.8\columnwidth]{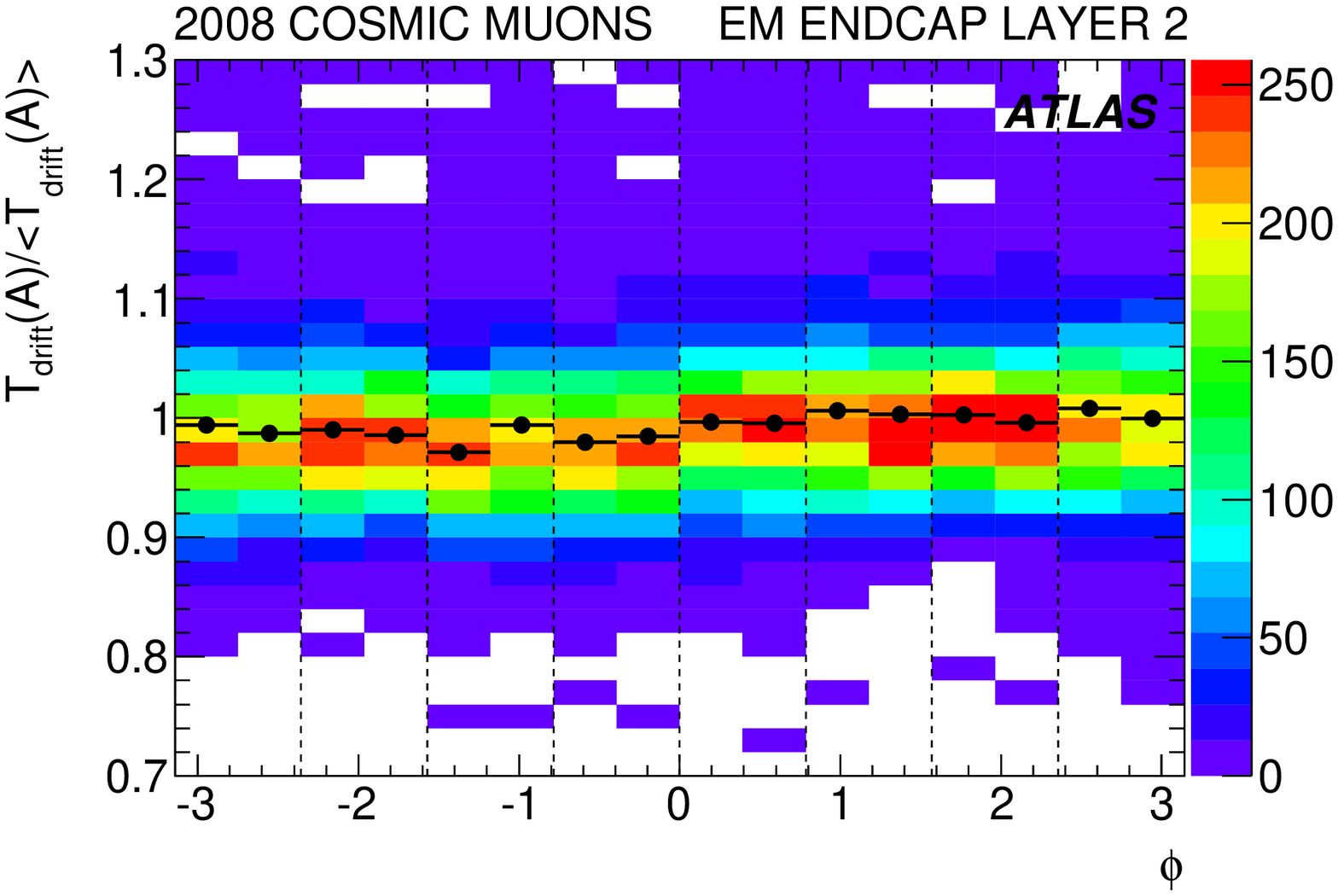}
}
\subfigure[layer 2 of endcap C]{
\label{F:tdriftvsphi_EMECC}
\includegraphics[width=0.8\columnwidth]{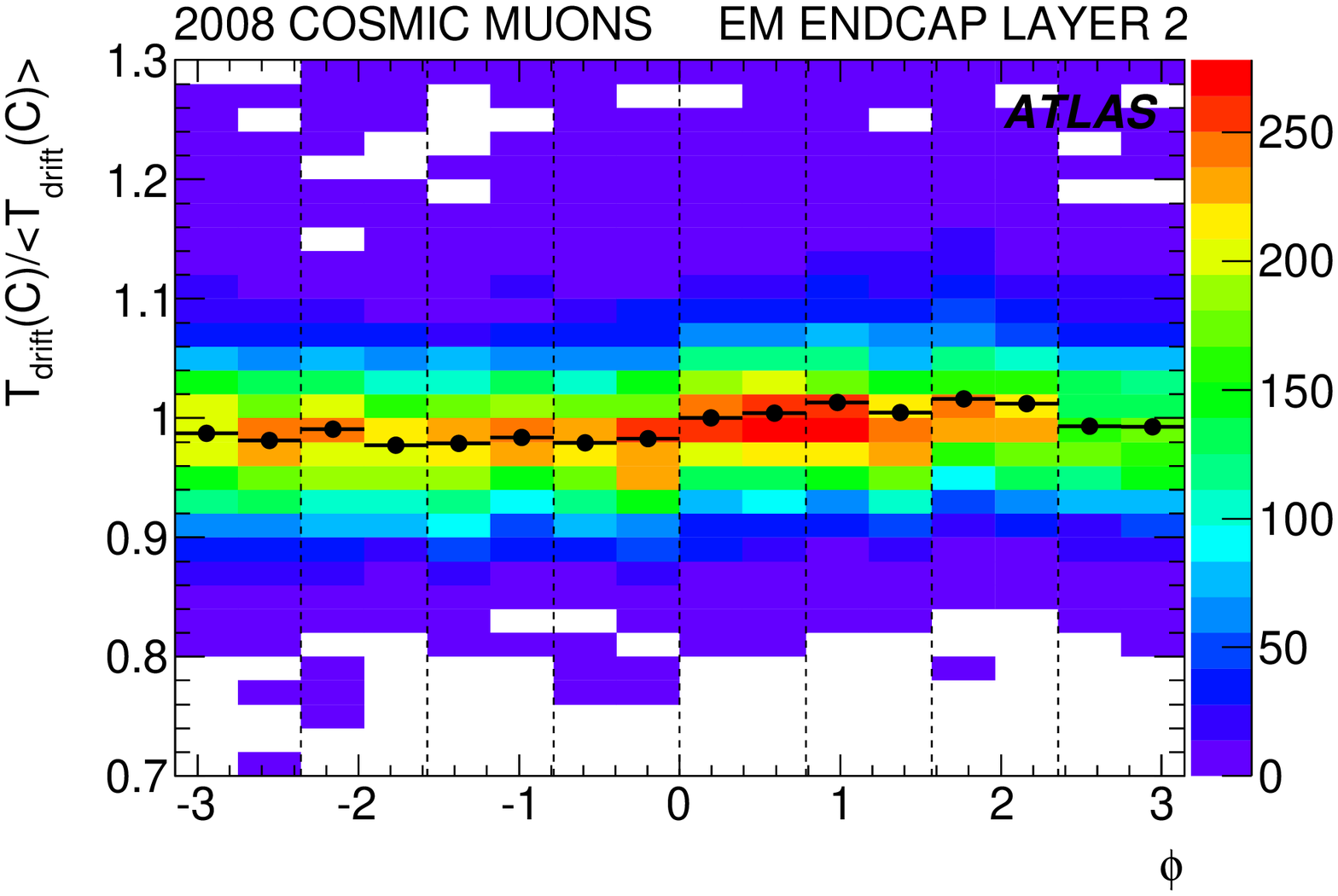}
}
\caption{Drift time normalized to the average value versus $\phi$ for layer 2 
of the $\eta>0$ (a) and $\eta<0$ (b) endcap wheels. The black dots are the 
average per $\phi$ bin and the vertical dashed lines show the boundaries 
between different modules.}
\label{F:TdriftvsPhi_EMEC}
\end{figure*}

Figure~\ref{F:TdriftvsPhi_EMEC} shows the drift time $T_{drift}$ as 
a function of azimuthal angle for layer 2 for the two endcaps. The 
values of $T_{drift}$ for each given pseudorapidity bin have been 
normalized to the average in order to mask the dependence on $\eta$.  
Vertical dashed lines indicate the boundaries between modules.
An asymmetry is visible on Figure~\ref{F:TdriftvsPhi_EMEC}
between positive and negative values of $\phi$: $T_{drift}(\phi > 0)$ is larger 
($0.996 \pm 0.002$) than $T_{drift}(\phi < 0)$ ($0.980 \pm 0.002$).
Since $\phi < 0$ is the lower half of the calorimeter, 
we associate this effect to the greater gravitational compression of
this part leading to slightly smaller gaps than in the upper
half $\phi > 0$. 

\subsection{Response uniformity}
\label{S:uni_EMEC}

An estimate of the intrinsic uniformity of the endcap can be made in a similar manner 
as presented for the barrel in Section~\ref{S:uni_EMB}. The average drift time across a region of
size $0.1 \times 0.1$ on the
($\eta$,$\phi$) plane is computed, with special care to take into account the varying gap thickness.

\begin{figure}[htbp]
\centering
{\includegraphics[width=0.8\columnwidth]{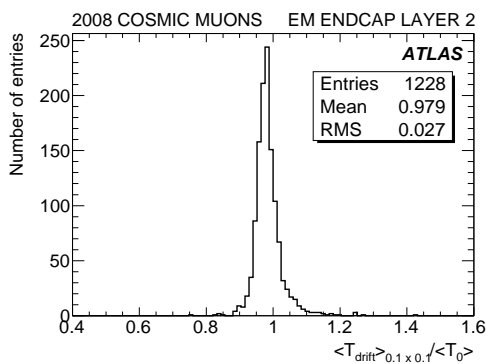}}
\caption{Drift time uniformity between groups of $4 \times 4$ cells 
  ($\Delta \eta \times \Delta \phi = 0.1 \times 0.1$) for endcap layer 2. The normalization 
  $<T_{0}>$ is obtained as a fit to the data using a first order polynomial in each HV 
  region to cancel out the influence of the gap variation with $\eta$.}
\label{F:uni_tdrift}
\end{figure}

Figure~\ref{F:uni_tdrift} 
represents the distribution of $T_{drift}/<T_{0}>$ for layer 2.
The normalization $<T_{0}>$ corresponds to the value (per $\eta$ cell) predicted from 
a first order polynomial fit to the data $T_{drift}$ in each high voltage region.
This normalization cancels out the change of the drift time due to the nominal design gap size variation with $\eta$.
The study is carried out only for layer 2 since it contains 
most of the shower energy of typical LHC electrons and photons.
In addition, more events have been recorded in layer 2 than in 
the other layers, which increases the statistical accuracy of the measurement.

The drift time uniformity of the $T_{drift}$ ($0.1 \times 0.1$)
distribution has an RMS of $(2.8\pm 0.1) \%$. To get the pure systematic 
non-uniformity between the $0.1 \times 0.1$ cells, the dispersion within the 
$0.1 \times 0.1$ cells, which in this case is not negligible, 
$(1.5 \pm 0.1) \ \%$, is quadratically subtracted. These numbers translate 
to a uniformity of the endcap calorimeter response due to intrinsic 
gap variations of $ (0.54 \pm 0.02) \%$. 
Systematic effects as discussed in Section~\ref{sec:systematics} and the uncertainty on
$\alpha$ (see Section~\ref{sect:ionization}) increase the error to $(0.54^{+0.06}_{-0.04}) \ \%$.

\subsection{Electrode shift}

\begin{figure}[htbp]
  \centering

    \includegraphics[width=0.8\columnwidth]{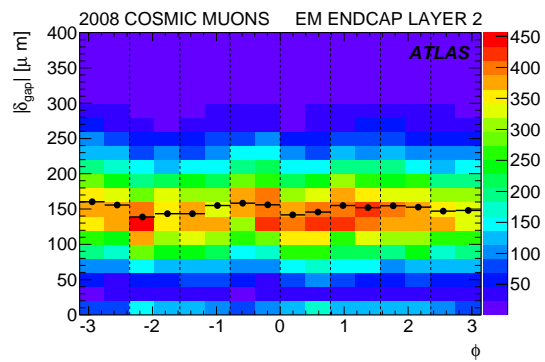}
 
  \caption{Electrode shift as function of $\phi$ for layer 2 of the endcap.
  The black dots are the average per $\phi$ bin and the vertical dashed lines show 
  the boundaries between different modules.}
  \label{F:shiftvsPhi}
\end{figure}

The distribution of the electrode shift as a function of the azimuthal angle is presented in
Figure~\ref{F:shiftvsPhi} for layer 2.  A rather flat behavior is observed.  Vertical dashed
lines correspond to the boundaries between consecutive modules. With a finer binning no
particular increase of the shift is observed at these transitions, even when extending the scale 
to $1000 \rm \mum$.  The average of about $146\rm m$ is independent of the layer.

\section{Drift time and velocity measurements}
\label{sect:driftvelo}

\begin{figure}[htbp]
\centering

\subfigure[Drift time]{
\label{F:tdrift_MiddALL}
\includegraphics[width=0.8\columnwidth]{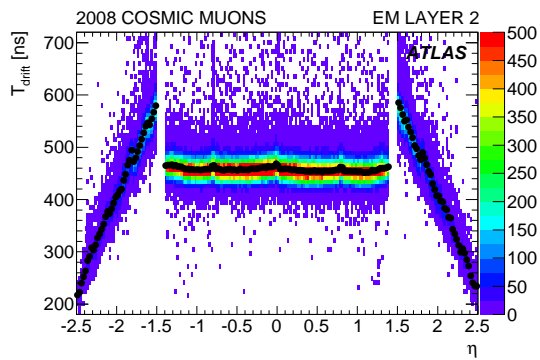}
}
\subfigure[Drift velocity]{
\label{F:vdrift_MiddALL}
\includegraphics[width=0.8\columnwidth]{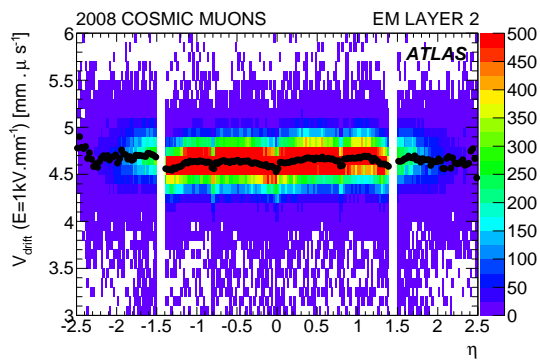}
}
\caption{(a) Drift time and (b) Drift velocity (at $E = 1 \rm \ kV/mm$) versus $\eta$
in layer 2. The black dots are the average per $\eta$ bin.}
\label{fig:results_EMEC}
\end{figure}

To quantify the consistency of the drift time measurements, the drift
velocity ($V_{drift}$) is studied more closely. The drift velocity can be extracted from drift time measurements
if the local gap values are accurately known (see Equation~\ref{eq:pure_vdrift}).
Both $w_{gap}$ and $T_{drift}$ are designed to be constant for the barrel,
but varying with pseudorapidity for the endcap. The variation of the
drift time $T_{drift}$ (see Figure~\ref{F:tdrift_MiddALL}) does not compensate for the variation of $w_{gap}$ because
$T_{drift} \sim w_{gap}^{1+\alpha}$. 
In addition, the different high voltage regions in the endcap introduce steps
in the behavior of the drift velocity as a function of $\eta$.

In order to compare accurately the drift velocities between barrel and endcap and for each
calorimeter layer, they are scaled to a reference field of $1 \rm \ kV/mm$:

\begin{equation}
V_{drift}(1\kV/\mm) = \frac{\it{w_{gap}}} {\it{T}_{\it{drift}}} \left( \frac { 2000\V \cdot \it{w_{gap}} } { \it{HV}_{\it{nom}} \cdot  \rm{2}\mm}\right) ^{\it{\alpha}}
\label{eq:velo1kv}
\end{equation} 

\noindent where $HV_{nom}$ is the nominal high voltage value, $w_{gap}$ is taken from the design value 
and $\alpha$ is the exponent introduced in Section~\ref{sect:ionization}.
Figure~\ref{F:vdrift_MiddALL} shows the drift velocity at the same field $1 \rm \ kV/mm$ for layer 2 of the entire calorimeter as a function of $\eta$.
As expected, a rather constant behavior is observed over the entire calorimeter.
The deviations from a perfect horizontal line is explained by local non-uniformities.
Deviations are observed at the transition regions at $\eta$=0 and $|\eta|=0.8$ and in the crack region between barrel and 
endcap at $|\eta|=1.4$, where the field is lower.

The temperature in the endcap A ($\eta>0$) is slightly higher (by about $0.3 \rm \ K$) than the temperatures of 
the barrel ($88.5 \rm \ K$) and endcap C ($88.4 \rm \ K$). 
This can explain the larger drift velocity measured in endcap C ($\eta<0$) with respect to endcap A, by
$\sim 0.6\%$ (see Figure~\ref{F:vdrift_MiddALL}), the expected difference being approximately $0.5\%$.

\begin{table*} [htbp]
  \centering
  \begin{tabular}{lll}
    \hline \hline
 & Layer &  Drift velocity (in $\rm mm/\mus$ at $1\rm \ kV/mm$) \\
\hline
\multirow{4}{*}{Barrel}
& Presampler & $4.52 \pm 0.001$ (stat) $^{+0.11}_{-0.07}$ (syst) \\
& Layer 1      & $4.62 \pm 0.003$ (stat) $^{+0.06}_{-0.14}$ (syst) \\
& Layer 2     & $4.63 \pm 0.002$ (stat) $^{+0.06}_{-0.14}$ (syst) \\
& Layer 3      & $4.59 \pm 0.002$ (stat) $^{+0.06}_{-0.14}$ (syst) \\
\hline
\multirow{3}{*}{Endcap}
& Layer 1      & $4.65 \pm 0.002$ (stat) $^{+0.10}_{-0.14}$ (syst) \\
& Layer 2     & $4.69 \pm 0.001$ (stat) $^{+0.10}_{-0.14}$ (syst) \\
& Layer 3       & $4.59 \pm 0.002$ (stat) $^{+0.10}_{-0.14}$ (syst) \\
    \hline \hline
  \end{tabular}
  \caption{Drift velocity at $E = 1 \rm \ kV/\mm$ in the different layers of the calorimeter.}
  \label{table:syst_number}
\end{table*}

\begin{figure}[htbp]
\centering

\subfigure[Barrel]{
\label{F:vdrift_allEMB}
\includegraphics[width=0.8\columnwidth]{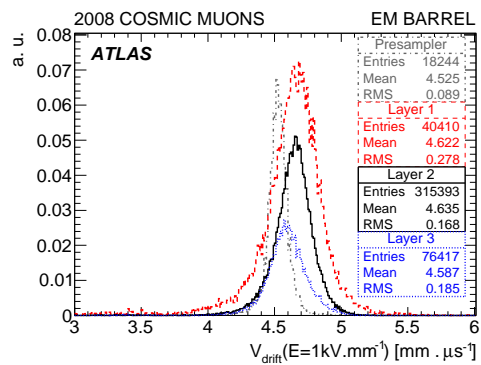}
}
\subfigure[Endcap]{
\label{F:vdrift_allEMEC}
\includegraphics[width=0.8\columnwidth]{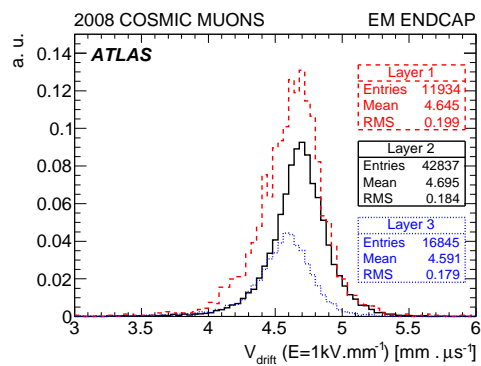}
}
\caption{Drift velocity distribution for the barrel (a) and endcap (b). }
\label{fig:vdrift_all}
\end{figure}

Figure~\ref{fig:vdrift_all} shows the comparison of $V_{drift}$ for the different layers of the barrel  
and endcaps. The mean values of the distributions are also quoted. The errors on these means, given the large
number of pulses averaged and the random nature of the noise dominating the error on single measurements,
are much smaller than the systematic uncertainties (see Section~\ref{sec:systematics}).  According to
Equation~\ref{eq:velo1kv}, the uncertainty in the drift velocity depends on
uncertainties in both the gap size and the drift time. The former can be
extracted from an azimuthal and pseudorapidity uniformity study, giving values smaller or equal to $1 \%$ 
and $2 \%$ for the barrel and endcap respectively.  The latter receives contributions from several
sources (see Section~\ref{sec:systematics}).
The  mean values of the drift velocity for the
different layers of the barrel and endcap
are given in Table~\ref{table:syst_number}.
They are all compatible within errors, although the barrel presampler 
is somewhat below the average. 

These results can be compared with the measurements from~\cite{VdvsTF}
which give $(4.65 \pm 0.12)\rm \ mm/\mus $ for 
a LAr temperature of $88.5\rm \ K$ and provides good agreement with the present measurement.


\section{Systematic uncertainties}
\label{sec:systematics}

The different sources of systematic uncertainties affecting the measurement
of the drift time which have been studied are discussed below.
The resulting systematic uncertainties on the velocity are given 
in Table~\ref{table:syst_number}, and in Sections~\ref{S:uni_EMB}
and~\ref{S:uni_EMEC} for what concerns the uniformity of response.

\subsection{Comparison of the results obtained in the barrel with the two prediction methods}

Two pulse shape prediction methods have been used for the barrel.
Their results are compared to give an estimate of the systematic uncertainty on the prediction.
For layers 2 and 3, the mean value of the difference between the predicted distributions is $\sim 0.2 \rm \ ns$ 
and the RMS in the $\eta$ direction is $\sim 1.2 \rm \ ns$  which is of the order of the
precision of the measurement for both methods: hence no significant difference
is observed for these layers. 
For layer 1, which also suffers from low statistics, the mean value of the difference 
($1.3 \rm \ ns$) (see Section~\ref{sect:bother}) is taken as an estimate of the systematic 
uncertainty associated with the prediction.

\subsection{Different fit strategies}

In addition to the fit procedure described in this paper, another approach was also followed in layer 2 of the barrel:
the cell-based fit. The method consists of fitting simultaneously all the ($N$) pulses 
collected in a given cell, using a single value for each of the drift time and the shift parameter, 
and $N$ global normalization factors and timing adjustments (one of each per pulse).
This yields results that are similar but not identical to those obtained from a weighted average of the 
individual fits with the weight $(S_{max}^{gain})^2$. For instance the average drift time in the
case of the cell-based fit is $1.2 \rm \ ns$ (i.e. $0.3 \%$) lower due to a somewhat reduced effect 
of pulses with large $T_{drift}$.
With the cell-based method, which has more statistical power for a given fit, 
the spike at zero visible in Figure~\ref{F:shift0} is very much reduced, 
confirming that it originates from statistical fluctuations of the
noise leading to a rising slope around $550 \rm \ ns$ steeper than for a 
single triangle.

\subsection{Variation of parameters of the cell}

The effect of the uncertainty on the capacitance in layer 2 of the barrel on the FPM 
determination of the drift time is studied as follows: the capacitance is varied by
 $\pm 5\%$ based on measurements, and a new set of the parameters $\tau_{sh}$ and $Z_S$
(defined in Section~\ref{sect:prediction}) are recalculated from the FPM calibration fits 
and used in new fits of the cosmic muon data. A small change in the overall drift time scale is observed,
but no significant variation in the drift time dependence
on $\eta$. It should nevertheless be noted that when varying the capacitance in either
direction, the drift times increased by about $3\ns$.
As discussed in~\cite{fpm}, an increase (decrease) in the value of the capacitance is
partially compensated by a smaller (larger) value of the shaper time constant
$\tau_{sh}$, which leads to only minor variations in the pulse shape.


For the RTM method, the estimated uncertainty for the determination of $LC$ and $\tau_{\rm cali}$ 
is less than $3\%$. The $\tau_{\rm cali}$ uncertainty induces an uncertainty of about $0.5 \%$ on $T_{drift}$, 
with an additional contribution of less than $0.1 \%$ coming from the $LC$ uncertainty.

\subsection{Effect of electron attachment}

In the presence of impurities in the LAr medium, drift electrons may attach to the impurities
with an associated lifetime $T_{live}$, and the signal shape is no longer triangular but has the form:
\begin{equation} 
I(t) = \frac{Q_0}{T_{drift}} e^{-t/T_{live}} \left(1-\frac{t}{T_{drift}}\right)
\hspace{0.1cm} {\rm for} \ \ 0<t<T_{drift}
\end{equation} 

Using the Fourier transform of  $I(t)$, the
pulse shape is derived by convolution of the various factors
affecting pulse formation and propagation (see \cite{fpm} for the
general case). The data are then fitted with the additional parameter $T_{live}$.

Although this new parametrization allows to reduce the size of residuals, the values
obtained for $T_{live}$ have a large dispersion 
 (about $6 \rm \ \mus$ for both the average and RMS of the distribution). Another
weak point of this description is that the effect is totally absent in the presampler,
which is in the same liquid bath as the calorimeter.

A systematic uncertainty of $^{+1.5}_{-0}\%$ in the drift time
is conservatively estimated from the difference between the cases of including or not the $T_{live}$ parameter.
The $\eta$ dependence of $T_{drift}$ remains essentially the same in both cases. 
The drift velocity remains unchanged in the presampler, but is reduced by $1.5$ to $2\%$ in
the other layers, which would make the presampler and the rest of the barrel more compatible.

While this study was made only in the barrel, the estimated systematic uncertainty is also
used for the endcap.

\subsection{Variation of the bent triangle contribution}

The amount of energy deposited in the bent sections of the calorimeter is estimated using the simulation. To
account for possible differences between data and simulation,  a systematic
uncertainty related to the estimate of the fraction of signal
collected in the bent sections is assessed by varying  the
contribution of the triangle associated with the bends $f_{bend}$ by $\pm$ $20 \ \%$
based on Table~\ref{tab:bend}. This test was done in a limited section of layer 2 of the barrel.
The resulting systematic variation of the drift time is $\mp 3 \rm \ ns$,
as if $T_{drift}$ were compensating the absence of the bent triangles.
It should be noted that the variations of the drift time with the relative amplitude of the
third and fourth triangle (see Equations~\ref{eq:td1} to~\ref{eq:td4}) are constant throughout the detector;
uncertainties on the contribution from bent sections should therefore not affect the
estimate of the intrinsic uniformity, except in layer 1 (see Section~\ref{sect:bother}). 

The procedure to estimate $T_{bend}$ and $f_{bend}$ in the endcap
requires that the contributions from the bent and straight parts of the accordion
can be separated using the local drift time distribution of simulated $10\rm \ GeV$ photon showers.
The uncertainty induced by this procedure is propagated to the
final $T_{drift}$ value, leading to a $0.2 \%$ variation that is compatible
with the precision of the measurement.

\subsection{Variation of the parameter $\alpha$}\label{sect:alpha}

The effect of the uncertainty on the exponent $\alpha$ in the determination of the
drift time in the endcap was studied by varying $\alpha$ in the range from $0.30$ to $0.39$
larger than the range determined in Section~\ref{sect:ionization} ($0.28$ to $0.34$). This
larger range was initially motivated by a previous measurement of this exponent during the beam test of the
endcap prototype using $120 \rm \ GeV$ electrons,  where a value of $0.39$ seemed to describe the data
better, however over a larger electrical field range than relevant here.
The effect of this difference ($0.30$ to $0.39$) on the drift time is approximately $1 \rm \ ns$ or about $0.2 \%$, 
again at the level of precision of the measurement. The effect on the retained range ($0.28$ to $0.34$) would
be even smaller. The arithmetic effect of the uncertainty on $\alpha$ on the uniformity was considered in
Sections~\ref{S:uni_EMB} and~\ref{S:uni_EMEC}.

\subsection{Summary of the systematic uncertainties}
The systematic uncertainties discussed above apply to the drift time measurements and can be translated
into drift velocity through the Equation~\ref{eq:pure_vdrift}. The drift velocities for each layer are 
summarized in Table~\ref{table:syst_number}.

Averaging over the presampler and layer 2 (barrel and endcap) values, for which most of the 
systematics are uncorrelated, gives as the final result for the reference field of $1 \rm \ kV/mm$ and 
a temperature $T = 88.5 \rm \ K$:

\begin{equation}
V_{ref} = (4.61 \pm 0.07) \rm \ mm/\mus 
\end{equation}

\section{Direct determination of local gap and drift velocity at operating point}
\label{sect:Vdrift_gap}

Taking advantage of the studies presented above, a somewhat more global treatment of the data is 
presented below, which allows:
\begin{itemize}
\item to unify the comparison of the local measured gaps, and their reference value from construction 
in both the barrel and the endcaps.
\item to obtain for the whole calorimeter the values of the drift velocity at the local operating points.
\end{itemize}

If the drift velocity were to be fully saturated, i.e. independent of the electric field, a measurement
of the drift time would trivially give the associated local gap using Equation~\ref{eq:pure_vdrift}.
In the situation analyzed here, the drift velocity depends weakly on the electric field,
with a power law already given in Section~\ref{sect:ionization} (see Equation~\ref{eq:vdrift_field}).

Using Equations~\ref{eq:pure_vdrift} and~\ref{eq:vdrift_field} rewritten below as

\begin{equation}
V_{drift} = V_{ref} \cdot \Big[ \frac{HV}{HV_{0}} \cdot \frac{w_{gap0}}{w_{gap}}\Big]^{\alpha}
\label{eq:vdrift_op}
\end{equation}

\noindent it is possible to express both the local velocity and the local gap, as functions of the measured $T_{drift}$:

\begin{equation}
w_{gap} = [A \cdot T_{drift}]^{1/(1+\alpha)}
\label{Eq:gap}
\end{equation}

\begin{equation}
V_{drift} = \frac{A^{1/(1+\alpha)}}{T_{drift}^{\alpha/(1+\alpha)}}
\label{Eq:vd}
\end{equation}

\noindent with $A = V_{ref} \cdot \Big[ \frac{HV}{HV_{0}}\Big]^{\alpha} \cdot w^{\alpha}_{gap0}$. The analysis presented
below uses: $\alpha=0.3$, $w_{gap0}=2 \rm \ mm$, $HV_{0}=2 \rm \ kV$ and normalizes the
drift velocity at $1 \rm \ kV/mm$ to the average $V_{ref}= 4.61 \rm \ mm/\mu s$, as reported in 
Section~\ref{sec:systematics}. The effect of the shift ($x \sim 0.1$) was estimated to bias the above analysis
  by less than $0.2 \ \%$ on the extracted gap value, and is therefore neglected. Data for the endcaps have been corrected for the temperature difference, and rescaled to $88.5 \rm \ K$.

\begin{figure}[htbp]
  \begin{center}
    \includegraphics[width=0.80\columnwidth]{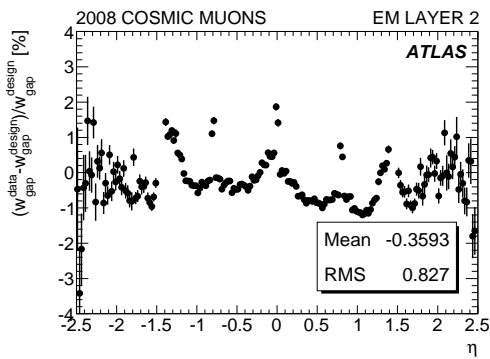}
  \end{center}
  \caption{Relative difference between the design gap values and the values extracted from $T_{drift}$ measurements.
  \label{F:Diffgap}}
\end{figure}

The additional information yielded by this analysis shows directly how the ratio of the
measured gap to the designed gap varies as a function of position in the detector. Figure~\ref{F:Diffgap} 
shows the relative difference between the calculated and design values. The average difference is not exactly $0$. 
This comes from the fact that the average velocity value used for the normalization 
includes presampler data, while the gap calculation presented in Figure~\ref{F:Diffgap}
contains only layer 2.

One can see that the ratio between calculated and design values, spanning a gap range between $1 \mm$
and $2.5 \mm$, has an RMS of $0.83 \%$, i.e. typically $16 \rm \ \mu m$. In the presampler, the corresponding 
dispersion is $7 \rm \ \mu m$, reflecting a more rigid fixing of the electrodes defining the gaps. 
In the barrel part one recognizes the systematic effects in the results discussed in Section~\ref{sect:bmid} 
(see in particular Figure~\ref{F:td_barmi_eta}) 
associated with the slight bulging of the absorbers, and the ``transition regions" at $\eta=0$, $\pm 0.8$ and $\pm 1.4$. 
Strictly speaking these transitions areas, for which additional effects enter into play, should be corrected for in the
calculation of the RMS. In the endcaps the statistical power is unfortunately lower
giving rise to larger fluctuations, but no significant trend is observed.

\begin{figure}[htbp]
  \begin{center}
    \includegraphics[width=0.80\columnwidth]{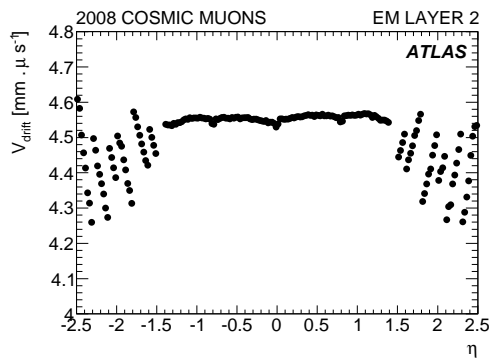}
  \end{center}
  \caption{Drift velocity versus $\eta$ in the layer 2 at the operating point extracted from $T_{drift}$ measurements.
  \label{F:velo_op}}
\end{figure}

Figure~\ref{F:velo_op} shows the drift velocity obtained using Equation~\ref{Eq:vd} 
as a function of pseudorapidity and the same normalization as above. 
As opposed to Figure~\ref{fig:results_EMEC},
which gave the velocity at a reference field of $1 \rm \ kV/mm$, Figure~\ref{F:velo_op} shows the drift
velocity at the local operating field, which is directly related to the peak current (see Section~\ref{sect:ionization}) associated
with an energy deposition.

In the barrel region, the drift velocity is essentially flat, with a slight modulation reflecting
the variation of the absorber thickness with pseudorapidity. Taking 
the average value of the drift velocity in sectors of $\Delta \eta \times \Delta \phi = 0.1\times 0.1$, 
as was done in Section~\ref{S:uni_EMB} for $T_{drift}$, one obtains a distribution with an RMS
 of $0.29\%$ exactly equal to what was derived in Section~\ref{S:uni_EMB} from the RMS of the $T_{drift}$ distribution,
 showing the expected consistency of the analyses using $T_{drift}$ or $V_{drift}$.
   
In the endcap region, one observes the 6 sawteeth on each side resulting from the finite granularity of
the HV distribution (see Figure~\ref{F:hv_distri}). Corrections are made in the energy reconstruction to normalize the response
of each strip in pseudorapidity to the response of the strip in the center of the HV sector, using 
the power law dependence. Beside these modulations, one observes that:

\begin{itemize}
\item the average velocity in the endcaps is smaller than in the barrel. In the energy reconstruction this is 
accounted for by correction factors (which also take into account the fact that the lead 
thicknesses are different) determined from test beam and implemented in the detailed Monte Carlo
simulation of the full ATLAS detector.

\item the measured velocity averaged over an HV sector somewhat diminishes with increasing pseudorapidity. This
effect goes in the same direction (lowering the response) as the reduced contribution of liquid argon
to showering/conversion effects at large pseudorapidities (small gaps). Both effects are qualitatively 
counterbalanced by the fact that the relative contribution of bends as compared to flat parts is lower at high pseudorapidity,
resulting in an increased response. As already mentioned, detailed Monte Carlo simulations normalized with test beam scans 
were used to determine the HV values optimizing the uniformity of response of the endcaps. This will be cross
checked when enough $Z^{0} \rightarrow e^{+}e^{-}$ decays become available.
\end{itemize}


\section{Conclusions}
\label{sect:conclusions}

We have shown in this paper that sufficient amounts of ionization data
($\sim 0.5$ million pulses of energy larger than $\sim 1 \rm \ GeV$) 
can be used for a precision measurement
of the average electron drift time in each cell of the highly granular LAr electromagnetic calorimeter of ATLAS
that has been readout with fast electronics, in the current mode.
In this regime, the recorded energy is directly proportional to the drift
velocity of ionization electrons, which is readily obtained from the drift time measurement.
Furthermore, the drift velocity and thus the recorded energy are $\sim 4$ times less sensitive to gap
variations than the drift time.

Taking advantage of these facts, we derived an estimate of the calorimeter non-uniformity 
of response due to gap size variations, of $(0.29^{+0.05}_{-0.04}) \ \%$ and $(0.54^{+0.06}_{-0.04}) \ \%$
respectively for the barrel and the
endcaps. The other main contribution to the intrinsic non-uniformity of the calorimeter is the 
dispersion of the thickness of the lead absorbers 
which contributes $0.18 \%$ for both barrel and endcaps~\cite{construction,construction_Endcap}.

The drift time is also an input needed in order to reconstruct the signal amplitude by optimal filtering.
An examination of the tails of the drift time distributions singles out ``transition areas"
of the calorimeter, in both azimuthal or pseudorapidity angle, where the electrical field is lower than
average due to ``edge effects". Some modulations in the third layer of the barrel have
also been observed.

The analysis method used to derive the drift time provides as another parameter the average 
absolute value of the amount the electrodes are off center between their two neighboring absorbers.
The values obtained are around $146 \rm \ \mum$ for both barrel and endcap accordion
layers, and are substantially smaller for the presampler ($66.5 \rm \ \mum$) as expected from its design.
   
The drift velocity, rescaled to a field of  $1 \rm \ kV/\mm$, is obtained from the drift time measurements 
leading to an average of $(4.61 \pm 0.07) \rm \ mm/\mus$. This value is compatible with previously published measurements 
at the same operating temperature of $88.5 \rm \ K$.

The measurements presented in this paper illustrate the
accuracy achieved with this method even using cosmic muon data, thus
demonstrating that it can be used to correct for the measured
gap variations in order to eventually reduce the constant term of the energy resolution, especially if
the measurements are repeated with collision data. It is
therefore important, in the quest to improve the energy resolution
constant term, that in the future these measurements be done with LHC
collision data.

\begin{acknowledgement}

\section*{Acknowledgments}

We are greatly indebted to all CERN's departments and to the LHC 
project for their immense efforts not only in building the LHC, 
but also for their direct contributions to the construction 
and installation of the ATLAS detector and its infrastructure. 
We acknowledge equally warmly all our technical colleagues in the 
collaborating institutions without whom the ATLAS detector could 
not have been built. Furthermore we are grateful to all the funding 
agencies which supported generously the construction and the 
commissioning of the ATLAS detector and also provided the computing 
infrastructure.

The ATLAS detector design and construction has taken about fifteen years, 
and our thoughts are with all our colleagues who sadly could not see its 
final realisation.

We acknowledge the support of ANPCyT, Argentina; Yerevan Physics 
Institute, Armenia; ARC and DEST, Australia; Bundesministerium 
$ \rm f\ddot{u}r$ Wissenschaft und Forschung, Austria; National Academy of
Sciences of Azerbaijan; State Committee on Science \& Technologies of 
the Republic of Belarus; CNPq and FINEP, Brazil; NSERC, NRC, and CFI, 
Canada; CERN; NSFC, China; Ministry of Education, Youth and Sports 
of the Czech Republic, Ministry of Industry and Trade of the Czech 
Republic, and Committee for Collaboration of the Czech Republic with CERN; 
Danish Natural Science Research Council; European Commission, through the 
ARTEMIS Research Training Network; IN2P3-CNRS and Dapnia-CEA, France; 
Georgian Academy of Sciences; BMBF, DESY, DFG and MPG, Germany; 
Ministry of Education and Religion, through the EPEAEK program PYTHAGORAS II 
and GSRT, Greece; ISF, MINERVA, GIF, DIP, and Benoziyo Center, Israel; 
INFN, Italy; MEXT, Japan; CNRST, Morocco; FOM and NWO, Netherlands; 
The Research Council of Norway; Ministry of Science and Higher Education, 
Poland; GRICES and FCT, Portugal; Ministry of Education and Research, 
Romania; Ministry of Education and Science of the Russian Federation, 
Russian Federal Agency of Science and Innovations, and Russian Federal 
Agency of Atomic Energy; JINR; Ministry of Science, Serbia; Department 
of International Science and Technology Cooperation, Ministry of Education 
of the Slovak Republic; Slovenian Research Agency, Ministry of Higher Education, 
Science and Technology, Slovenia; Ministerio de Educaci\'on y Ciencia, Spain; 
The Swedish Research Council, The Knut and Alice Wallenberg Foundation, Sweden; 
State Secretariat for Education and Science, Swiss National Science Foundation, 
and Cantons of Bern and Geneva, Switzerland; National Science Council, 
Taiwan; TAEK, Turkey; The Science and Technology Facilities Council and 
The Leverhulme Trust, United Kingdom; DOE and NSF, United States of America.
\end{acknowledgement}


\bibliographystyle{atlasnote}
\bibliography{drifttime}


\end{document}